\documentclass[aps,pra,pdf,superscriptaddress,twocolumn,showpacs,nofootinbib]{revtex4-2}
\usepackage{amsmath,amssymb,amsfonts}
\usepackage{eqnarray}
\usepackage{graphics,float}
\usepackage{bm}
\usepackage{braket}
\usepackage{amsmath}
\usepackage{physics}
\usepackage{microtype}
\usepackage{mathtools}
\usepackage{color,xcolor,colortbl}
\usepackage{tikz}
\usetikzlibrary{shapes}

\usepackage[colorlinks=true,
linkcolor=blue,
filecolor=magenta,      
urlcolor=blue,
citecolor=blue]{hyperref}

\definecolor{Gray}{gray}{0.85}
\definecolor{LightCyan}{rgb}{0.25,0.9,0.95}
\definecolor{LightMagenta}{rgb}{0.91,0.21,0.96}
\definecolor{lightgreen}{rgb}{0.46,0.98,0.01}
\definecolor{lightviolet}{rgb}{0.61,0.37,0.92}
\newcommand{\marker}[2]{\raisebox{0.5pt}{\tikz{\node[draw,scale=0.65,#2,color=#1,fill=#1](){};}}}
\newcommand{\markersquare}[2]{\raisebox{0.5pt}{\tikz{\node[draw,scale=0.8,#2,color=#1,fill=#1](){};}}}
\newcommand{\markertriangle}[1]{\raisebox{0.5pt}{\tikz{\node[draw,scale=0.45,regular polygon, regular polygon sides=3,color=#1,fill=#1](){};}}}
\newcommand{\markerpentagon}[1]{\raisebox{0.5pt}{\tikz{\node[draw,scale=0.65,regular polygon, regular polygon sides=5,color=#1,fill=#1](){};}}}

\graphicspath{{Plots/}}

\begin{document}
	\title{Induced supersolidity in a Dy-Er mixture}
	\author{Soumyadeep Halder}
	\email{soumya.hhs@gmail.com}
	\affiliation{Department of Physics, Indian Institute of Technology Kharagpur, Kharagpur 721302, India}
	
	\author{Subrata Das}
	\email{subrata@vt.edu}
	\affiliation{Department of Physics, Indian Institute of Technology Kharagpur, Kharagpur 721302, India}
    \affiliation{Department of Physics, Virginia Tech, Blacksburg, Virginia 24061, USA}
	
    \author{Sonjoy Majumder}
	\email{sonjoym@phy.iitkgp.ac.in}
	\affiliation{Department of Physics, Indian Institute of Technology Kharagpur, Kharagpur 721302, India}
	
	\date{\today}
\begin{abstract}
Recent experimental realization of the heteronuclear dipolar mixture of Dy and Er atoms opens fascinating prospects for creating intriguing novel phases in dipolar quantum gases. The experimentally measured value of intra-species $s$-wave scattering length of $^{166}$Er condensate in a $^{164}$Dy-$^{166}$Er mixture is larger than its intra-species dipolar length, implies that the $^{166}$Er condensate itself will not be in a regime of dominated dipole-dipole interaction (DDI). However, we find that the presence of $^{164}$Dy atoms with high magnetic moment induces droplet nucleation and supersolidity in $^{166}$Er condensate via the long-range and anisotropic inter-species DDI. Remarkably, we find that the imbalance in the magnetic dipole moment combined with its strong anisotropic coupling led to the emergence of unique ground state phases. The emerging phases include doubly superfluid states, a mixture of insulating droplets and supersolid states, binary supersolids with uniform and alternating domains and a combination of supersolid-superfluid mixed states. We delineate the properties of all these ground state phases and construct a phase diagram. We also explore the dynamical evolution across these phase boundaries via a linear quench of inter-species scattering length. Although we have demonstrated the result for the $^{164}$Dy-$^{166}$Er mixture, our results are generally valid for other dipolar bosonic mixtures of different Dy-Er isotope combinations and may become an important benchmark for future experimental scenarios.
\end{abstract} 
	
\maketitle

\section{Introduction}
Supersolid (SS) is an intriguing state of matter that combines superfluidity and a spatial crystalline structure. This unique state was predicted nearly six decades ago and initially searched for in $^4$He \cite{gross_1957_unified, leggett_1970_can, toennies_2001_superfluid, toennies_2004_superfluid, barranco_2006_helium, boninsegni_2012_colloquium, chan_2013_overview, ancilotto_2017_density, kim_2004_probable, kim_2012_absence}. Despite being elusive in solid Helium, recent advancements in ultracold quantum gasses have made it possible to realize SS features. Bose-Einstein condensates (BEC) of Rydberg atoms \cite{cinti_2010_supersolid, henkel_2010_threedimensional, henkel_2012_supersolid}, with spin-orbit coupling \cite{wang_2010_spinorbit, li_2013_superstripes, li_2016_spinorbit, li_2017_stripe, bersano_2019_experimental, putra_2020_spatial, sachdeva_2020_selfbound, geier_2021_exciting}, and in optical cavities \cite{leonard_2017_supersolid, zhang_2022_supersolid} have already demonstrated the formation of SS-like phases. The realization of the SS phase in ultracold dipolar bosonic gases composed of highly magnetic rare-earth Dy \cite{ferrier-barbut_2016_observation, kadau_2016_observing, schmitt_2016_selfbound, ilzhofer_2021_phase, tang_2015_wave, tang_2015_bose} and Er \cite{chomaz_2016_quantumfluctuationdriven, chomaz_2019_longlived, natale_2019_excitation} atoms attract significant attention in recent years. In dipolar bosonic gases, the mean-field-driven collapse is arrested by quantum fluctuations \cite{lima_2011_quantum, lima_2012_meanfield, petrov_2015_quantum, chomaz_2016_quantumfluctuationdriven}. The interplay between the short-range isotropic contact interaction (CI) and long-range anisotropic dipole-dipole interaction (DDI) in a dipolar BEC gives rise to a rich plethora of physical phenomena including the anisotropic superfluidity \cite{mulkerin_2013_anisotropic, martin_2017_vortices}, the appearance of roton excitation \cite{santos_2003_rotonmaxon, bisset_2013_roton, baillie_2017_collective, chomaz_2018_observation, schmidt_2021_roton}, formation of self-bound droplet \cite{wachtler_2016_quantum, wachtler_2016_groundstate, baillie_2018_droplet, mishra_2020_selfbound, ghosh_2022_droplet} and SS phases both in quasi one \cite{bottcher_2019_transient, tanzi_2019_observation, roccuzzo_2019_supersolid, blakie_2020_supersolidity, smith_2023_supersolidity, sanchez-baena_2023_heating} and two-dimensional \cite{norcia_2021_twodimensional, bland_2022_twodimensional} trapping confinement. Theoretical and experimental studies over the past few years reveal exotic pattern formation in dipolar BEC \cite{hertkorn_2021_pattern, zhang_2021_phases, poli_2021_maintaining, gallemi_2022_superfluid, arazo_2023_selfbound, chomaz_2022_dipolar, sohmen_2021_birth}, nucleation of quantized vortices in SS state \cite{prasad_2019_vortex, roccuzzo_2020_rotating, gallemi_2020_quantized, sindik_2022_creation, klaus_2022_observationa}, formation of self-bound thin disk-shape macro droplets \cite{halder_2022_controla} and SS state in the anti-dipolar regime \cite{kirkby_2023_spin, mukherjee_2023_supersolid}.\par 

Binary mixture offers more exciting facets resulting from the interplay between intra- and inter-species interactions. Isotropic miscible quantum droplets have been realized in non-dipolar binary homonuclear \cite{semeghini_2018_selfbound, cheiney_2018_bright, ferioli_2019_collisions, flynn_2022_quantum} and heteronuclear \cite{guo_2021_leehuangyang, derrico_2019_observation} bosonic mixtures with attractive inter-species CI. Non-dipolar binary mixtures of alkali atoms can form polar molecules possessing a large electric dipole moment featuring various exotic supersolid states \cite{schmidt_2022_selfbound}. \par

Recent experimental realization of a quantum degenerate dipolar mixture of Dy and Er atoms \cite{trautmann_2018_dipolar}, and the ability to control their inter-species interaction strength through the Feshbach resonance \cite{durastante_2020_feshbach, politi_2022_interspecies} unveil new exciting aspects for the study of binary dipolar mixtures. Along with the intra- and inter-species CI, both components also exhibit long-range and anisotropic DDI, giving rise to the emergence of unique miscible and axially immiscible self-bound droplet states \cite{bisset_2021_quantum, smith_2021_quantum, smith_2021_approximate, lee_2021_miscibility}. Moreover, recent theoretical studies indicate that homonuclear binary dipolar mixtures have the potential to yield miscible and immiscible doubly SS and various mixed states both in quasi-one \cite{scheiermann_2023_catalyzation} and two-dimensional \cite{halder_2023_twodimensional} trap confinement. However, in such scenarios, the substantial DDI necessitates the quantum stabilization of both condensates through the Lee-Huang-Yang (LHY) correction mechanism. Furthermore, the relatively high peak densities in these droplet and SS phases shorten their lifetime. On the contrary, a dipolar-nondipolar mixture can sustain a long-lifetime supersolid state \cite{li_2022_longlifetime, bland_2022_alternatingdomain}. In the dipolar-nondipolar mixture, the dipolar component exhibits an insulating droplet (ID) or SS state, while the non-dipolar component serves as a background superfluid. However, the absence of inter-species DDI in such a system limits the physical richness and the emergence of the doubly SS phase is only feasible under the immiscible condition. \par

In contrast, both the condensates in a Dy-Er mixture possess substantial yet distinct magnetic dipole moments, resulting in a robust long-range anisotropic DDI coupling between the condensates. The imbalance in the DDI strength together with the strong intra- and inter-species interactions (CI and DDI), promises a more profound array of complex many-body phenomena. Driven by these facts, we uncover some of the diverse physical phenomena inherent in the Dy-Er mixture.\par 

In this article, we theoretically investigate the possibility of forming various intriguing ground state phases of the $^{164}$Dy-$^{166}$Er mixture confined in a quasi-two-dimensional trapping geometry. As per recent experimental findings, the intra-species $s$-wave scattering length of $^{166}$Er condensate in a $^{164}$Dy-$^{166}$Er mixture indicates that the $^{166}$Er condensate will not be in a dipole-dominated interaction regime. However, in this work, we have demonstrated that even within this experimentally constrained parameter regime, the $^{164}$Dy condensate can induce supersolidity in $^{166}$Er condensate. This intriguing phenomenon arises as a consequence of the anisotropic long-range DDI coupling between the two condensates. In our study, we observe fascinating ground state phases in the $^{164}$Dy-$^{166}$Er mixture, including a doubly superfluid (SF-SF), a mixed state featuring ID and SS, a uniform domain supersolid (UDS), an alternating domain supersolid (ADS) and a SS-SF mixed states. We characterize these distinct miscible and immiscible ground state phases by quantifying the overlap between the two species and determining the superfluid fraction of each species. We delineate all these phase boundaries and construct the captivating groundstate phase diagram within the experimentally feasible parameter range. Employing the time-dependent coupled extended Gross-Pitaevskii equation (eGPE), we also monitor the dynamical evolution of these phases during quench-induced dynamics. \par 

This paper is structured as follows. Section \ref{secii} describes the theory and formalism, including the coupled eGPE. Section \ref{seciii} is devoted to investigating all possible ground state phases of a $^{164}$Dy-$^{166}$Er mixture. In subsection \ref{seciiia}, we describe characterization processes to differentiate all possible ground state phases. In subsection \ref{seciiib}, we demark all these phase boundaries and construct the ground state phase diagram of a dipolar heteronuclear mixture of $^{164}$Dy and $^{166}$Er atoms. Subsection \ref{seciiic} is devoted to comprehending how the interplay between intra- and inter-species CI and DDI results in the formation of these intriguing ground state phases. In Sec. \ref{seciv}, we explore real-time dynamics and the formation of various two-dimensional miscible-immiscible droplet and supersolid states by using the time-dependent eGPE. A summary of our findings, together with future aspects, is provided in Sec. \ref{secv}. Appendix \ref{A} describes the ingredients of our numerical simulations. 
%%%%%%%%%%%%%%%%%%%%%%%%%%%%%%%%%%%%%%%%%%%%%%%%%%%%%%%%%%%%%%%%%%%%%%%%%%%%%%%%%%%%%%%%%%%
% Theory

\section{Formalism}\label{secii}
We consider a heteronuclear binary dipolar mixture consisting of $^{164}$Dy and $^{166}$Er atoms. Here we assume both species have the same atomic mass $m=165$u and neglect the effect of gravitational sag. This is a valid approximation since $^{164}$Dy and $^{166}$Er have a relative mass difference of less than 1.5\%. The atoms of the mixture are polarized along the $z$ direction by an external magnetic field and confined in a circular symmetric harmonic trapping potential. At zero temperature, the temporal evolution of the macroscopic wave function $\psi_i$ $(i=1,2)$ is described by the coupled extended GP equations $i\hbar\dot{\psi_i}=\mathcal{H}^{\rm GP}_{i}\psi_i$, with,
\begin{align}
	\mathcal{H}_i^{\rm GP}&=\Big[-\frac{\hbar^2}{2m}\nabla^2+V_t(\textbf{r})+\sum_{j=1}^{2}\Big(g_{ij}\abs{\psi_j(\bf{r},t)}^2+\nonumber\\&\int d\textbf{r}' V_{ij}^{\rm dd}(\textbf{r}-\textbf{r}')\abs{\psi_j(\bf{r}^{\prime},t)}^2\Big)+\Delta \mu_i\Big].\label{hami}
\end{align}
Here, $V_t(\textbf{r})=\frac{1}{2}m_i\omega^2(x^2+y^2+\lambda^2z^2)$ is the harmonic trapping potential with angular frequencies $\omega_x=\omega_y=\omega,\omega_z$ and $\lambda=\omega_z/\omega$ is the trap aspect ratio. The contact interaction strength between the atoms of species $i$ and $j$ is characterized by the coupling constant $g_{ij}=4\pi\hbar^2a_{ij}/m$ with $a_{ij}$ being the $s$-wave scattering length of the atoms. In addition to the contact interactions, there exists a long-range DDI between the atoms and it takes the following form
\begin{equation}
	V_{ij}^{\rm dd}(\textbf{r})=\frac{3g_{ij}^{\rm dd}}{4\pi}\left(\frac{1-3\cos^2\theta}{r^3}.\right),\label{dip_pot}
\end{equation}
Here $g_{ij}^{\rm dd}=4\pi\hbar^2a_{ij}^{\rm dd}/m$ is the dipolar coupling constant with the dipolar length $a_{ij}^{\rm dd}=\mu_0\mu_i^m\mu_j^m m/12\pi\hbar^2$, where $\mu_0$ is the vacuum permeability and $\theta$ is the angle between the axis linking two dipolar atoms and their dipole polarization direction ($z$-axis). The last term appearing in the Hamiltonian [Eq.({\ref{hami}})] represents the correction to the chemical potential resulting from the effect of quantum fluctuation (known as LHY correction) given by \cite{bisset_2021_quantum,smith_2021_approximate,smith_2021_quantum}
\begin{equation}
	\Delta \mu_i=\frac{m_i^{3/2}}{3\sqrt{2}\pi^2\hbar^3}\sum_{\pm}\int_{0}^{1}\dd u~\Re{I_{i\pm}},
\end{equation}
where
\begin{align}
	I_{1\pm}=\bigg(\tilde{U}_{11}\pm &\frac{\delta_1 \tilde{U}_{11}+2\tilde{U}_{12}^2 n_2}{\sqrt{\delta_1^2+4\tilde{U}_{12}^2n_1 n_2}}\bigg)\bigg(n_1\tilde{U}_{11}+\nonumber\\& n_2\tilde{U}_{22}\pm\sqrt{\delta_1^2+4\tilde{U}_{12}^2n_1 n_2}\bigg)^{3/2},
\end{align}
with $\delta_1=n_{1}\tilde{U}_{11}-n_2\tilde{U}_{22}$, and $\tilde{U}_{ij}(u)=g_{ij}[1+a_{ij}^{\rm dd}/a_{ij}(3u^2-1)]$, being the Fourier transform of the DDI potential. A similar expression for $\Delta\mu_2$ can be easily obtained with $\delta_2=-\delta_1$.\par 
 However the Dy and Er have magnetic dipole moments $\mu_1^m=9.93\mu_B$ and $\mu_2^m=7\mu_B$, which corresponds to the dipolar lengths $a_{11}^{\rm dd}=131a_B$ and $a_{22}^{\rm dd}=65.5a_B$, respectively. Whereas the experimentally measured\footnote[1]{To produce a $^{164}$Dy-$^{166}$Er mixture experimentally, a magnetic field of $B=2.075$ G \cite{trautmann_2018_dipolar} is required for the evaporative cooling process which corresponds to the value of $a_{11}=90a_B$ and $a_{22}=80a_B$ for $^{164}$Dy and $^{166}$Er condensate, respectively \cite{politi_2022_interspecies, tang_2015_wave, chomaz_2016_quantumfluctuationdriven}.} value of intra-species scattering lengths of $^{164}$Dy and $^{166}$Er condensates are $a_{11}=90a_B$ and $a_{22}=80a_B$, respectively \cite{trautmann_2018_dipolar, durastante_2020_feshbach, politi_2022_interspecies, tang_2015_wave, chomaz_2016_quantumfluctuationdriven}. Since $a_{22}>a_{22}^{\rm dd}$, it implies that in a $^{164}$Dy-$^{166}$Er mixture, $^{166}$Er condensate itself is not in a dipole-dominated interaction regime and stable against the mean field collapse. Therefore, the LHY correction is mainly relevant for the $^{164}$Dy condensate\footnote[2]{We observe that when we neglect the LHY correction for the $^{166}$Er condensate, the system's behavior remains unchanged, and it remains stable. However, in the absence of the LHY correction for the $^{164}$Dy condensate, the system will no longer be stable.}. Under this condition, the correction to the chemical potential energy is well approximated by the known form of LHY correction for a single species dipolar BEC $\Delta \mu_1=\gamma_1(\epsilon_{11}^{\rm dd}) n_1^{3/2}$ \cite{lima_2011_quantum, lima_2012_meanfield}, where the quantum fluctuation strength $\gamma_1 (\epsilon_{11}^{\rm dd})$ can be written as
\begin{equation}
	\gamma_1(\epsilon_{11}^{\rm dd})= \frac{32}{3}g_{11}\sqrt{\frac{a_{11}^3}{\pi}}F(\epsilon_{11}^{\rm dd}),
\end{equation}
with
\begin{equation}
	F(\epsilon_{11}^{\rm dd})= \Re{\int_0^1\dd u (1+\epsilon_{11}^{\rm dd}(3u^2-1))^{5/2}}
\end{equation}
and the dimensionless parameter $\epsilon_{11}^{\rm dd}=a_{11}^{\rm dd}/a_{11}$, quantifies the relative strength of DDI to the contact interaction between the atoms of $^{164}$Dy. The order parameters of each condensate are normalized to the total number of atoms in that species, $N_i=\int \dd\textbf{r}\abs{\psi_i(\textbf{r})}^2$.

%%%%%%%%%%%%%%%%%%%%%%%%%%%%%%%%%%%%%%%%%%%%%%%%%%%%%%%%%%%%%%%%%%%%%%%%%%%%%%

\section{Groundstate phases of \texorpdfstring{D\MakeLowercase{y}-E\MakeLowercase{r}}{Dy-Er} mixture}\label{seciii}
\begin{figure}[tb!]
	\centering
	\includegraphics[width=0.45\textwidth]{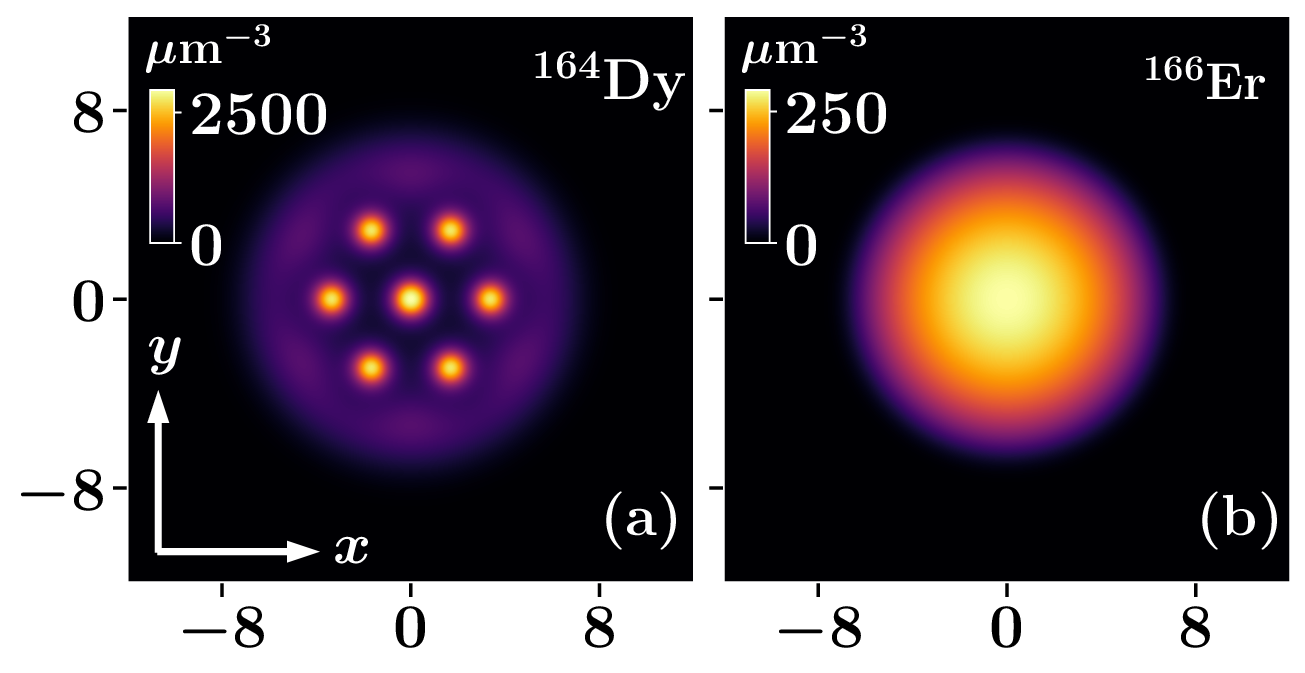}
	\caption{Ground state density profile of single component dipolar BEC. (a) Density distributions of Dy condensate with $\mu_1^m=9.93\mu_B$ and $a_{11}=90a_B$ and (b) Er condensate with $\mu_2^m=7\mu_B$ and $a_{22}=80a_B$ in the $x-y$ plane, where $a_B$ is the Bohr radius and $\mu_B$ is the Bohr Magneton. Each of these single component dipolar BEC consists of $N=6\times10^4$ number of atoms and is confined in a circular symmetric harmonic trap with $(\omega_x,\omega_y,\omega_z)=2\pi\times(45,45,133)$ Hz. The different color bar for each panel denotes the density of the corresponding condensate in units of $\mu\rm m^{-3}$.}\label{fig:1}
\end{figure}

In this section, we investigate the groundstate phases of a $^{164}$Dy-$^{166}$Er mixture with an equal number of atoms in each condensate ($N_1=N_2=N$) and the intra-species $s$-wave scattering lengths $a_{11}=90a_B (<a_{11}^{\rm dd}=131a_B)$ and $a_{22}=80a_B(>a_{22}^{\rm dd}=65.5a_B)$ \cite{tang_2015_wave, chomaz_2016_quantumfluctuationdriven}. We consider a pancake-shaped trap geometry to confine the mixture with $(\omega_{x},\omega_y,\omega_z)=2\pi\times(45,45,133)$ Hz. With this choice of parameters and without any inter-species coupling (single-component case) the Dy atoms form a periodic density modulated SS state, whereas the Er atoms form a SF state as shown in Fig. \ref{fig:1}. However, the presence of inter-species coupling in a Dy-Er mixture makes it a more intriguing candidate to investigate new possibilities in dipolar quantum gases. In the following, we evaluate the groundstate phases of the mixture as a function of the number of particles $N$ and inter-species scattering length $a_{12}$. We observe that the Dy condensate can induce droplet nucleation and supersolidity in the Er condensate.\\

 \subsection{Characterization of distinct ground state phases} \label{seciiia}
\begin{figure}[tb!]
	\centering
	\includegraphics[width=0.45\textwidth]{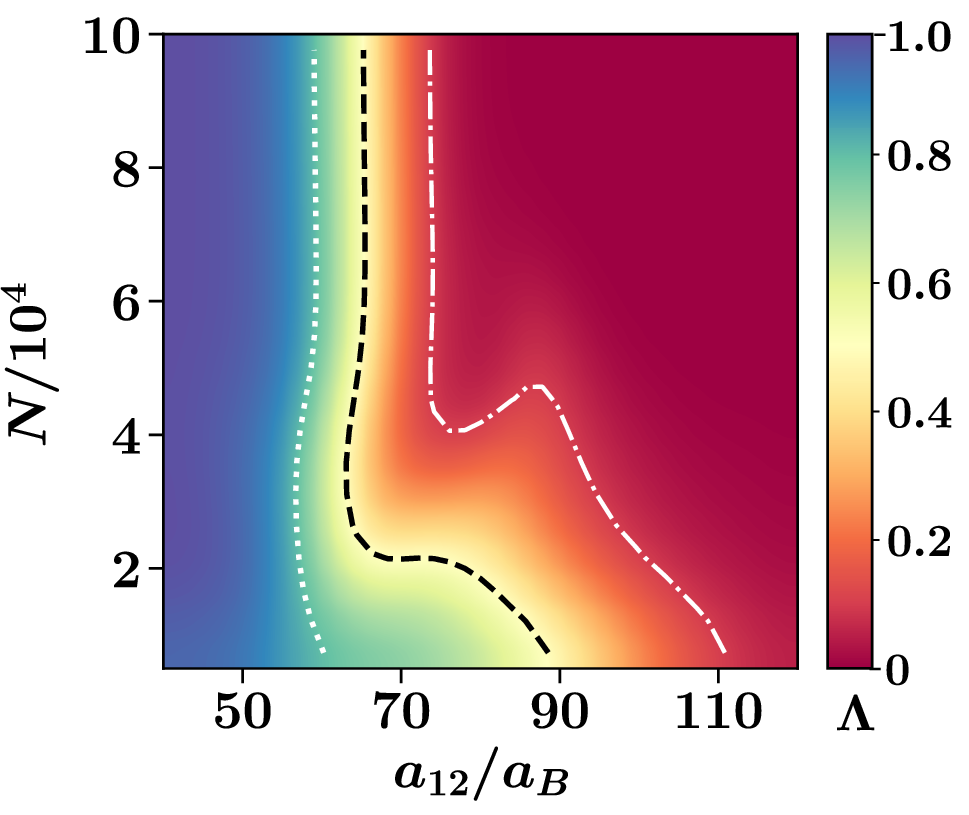}
	\caption{Value of the overlap integral $\Lambda$ of Dy-Er mixture as a function of inter-species scattering length $a_{12}$ and the number of particles in each condensate $N$. The black dashed contour has been drawn at $\Lambda=0.5$, roughly separating the miscible and immiscible phase domain. To demark the completely miscible and immiscible phase domain, two other dotted and dash-dotted white contours are drawn at $\Lambda=0.9$ and $\Lambda=0.1$, respectively. The color bar corresponds to the value of the overlap integral $\Lambda$. The result is for the case of binary dipolar BEC consisting of $^{164}$Dy and $^{166}$Er atoms with $a_{11}=90a_B$, $a_{11}^{\rm dd}=131a_B$ and $a_{22}=80a_B$, $a_{22}^{\rm dd}=65.5a_B$, respectively. The binary mixture is confined in a harmonic trap with $(\omega_x,\omega_y,\omega_z)=2\pi\times(45,45,133)$ Hz.}\label{fig:2}
\end{figure}

Dy-Er mixture can exhibit either a miscible or an immiscible phase. We differentiate these two phases by evaluating the overlap integral \cite{chen_2019_immiscible, bandyopadhyay_2017_dynamics}
\begin{equation}
    \Lambda=\frac{\left[\int d\textbf{r}n_1(\textbf{r})n_2(\textbf{r})\right]^2}{\left[\int d\textbf{r}n_1^2(\textbf{r})\right]\left[\int d\textbf{r}n_2^2(\textbf{r})\right]},\label{lambdao}
\end{equation}
where $n_i(\textbf{r})=\abs{\psi_i(\textbf{r})}^2$ is the densities of the species-$i$. $\Lambda=1$ implies maximal spatial overlap between the condensates, i.e., the system is in a completely miscible state, whereas a complete phase separation (immiscible phase) corresponds to $\Lambda=0$. The variation of overlap between the Dy and Er condensates with the number of particles $N$ and inter-species scattering length $a_{12}$ is depicted in Fig. \ref{fig:2}.\par 
The periodic density modulation due to the anisotropic DDI can be characterized by the density contrast \cite{smith_2023_supersolidity}
\begin{equation}
    \mathcal{C}=\frac{(n_i^{\rm max}-n_i^{\rm min})}{(n_i^{\rm max}+n_i^{\rm min})}.\label{contrast}
\end{equation}
Here $n_i^{\rm max}$ and $n_i^{\rm min}$ denote the neighboring maxima and minima of the $i$-th component observed on the x-y plane. A SF state is associated with a smooth density distribution where $n_i^{\rm max}= n_i^{\rm min}$, leading to $\mathcal{C}=0$. In case of an ID state where there is no overlap between the droplets ($n_{\rm min}\approx 0$), Eq. \ref{contrast} yields $\mathcal{C}\approx 1$. Conversely, a SS state is characterized by overlapping droplets ($n_i^{\rm min}\neq 0$), resulting in a density contrast $\mathcal{C}$ that attains an intermediate value between 0 and 1 \cite{halder_2023_twodimensional,halder_2022_controla}.\par 
\begin{figure}[tb!] 
	\centering
	\includegraphics[width=0.45\textwidth]{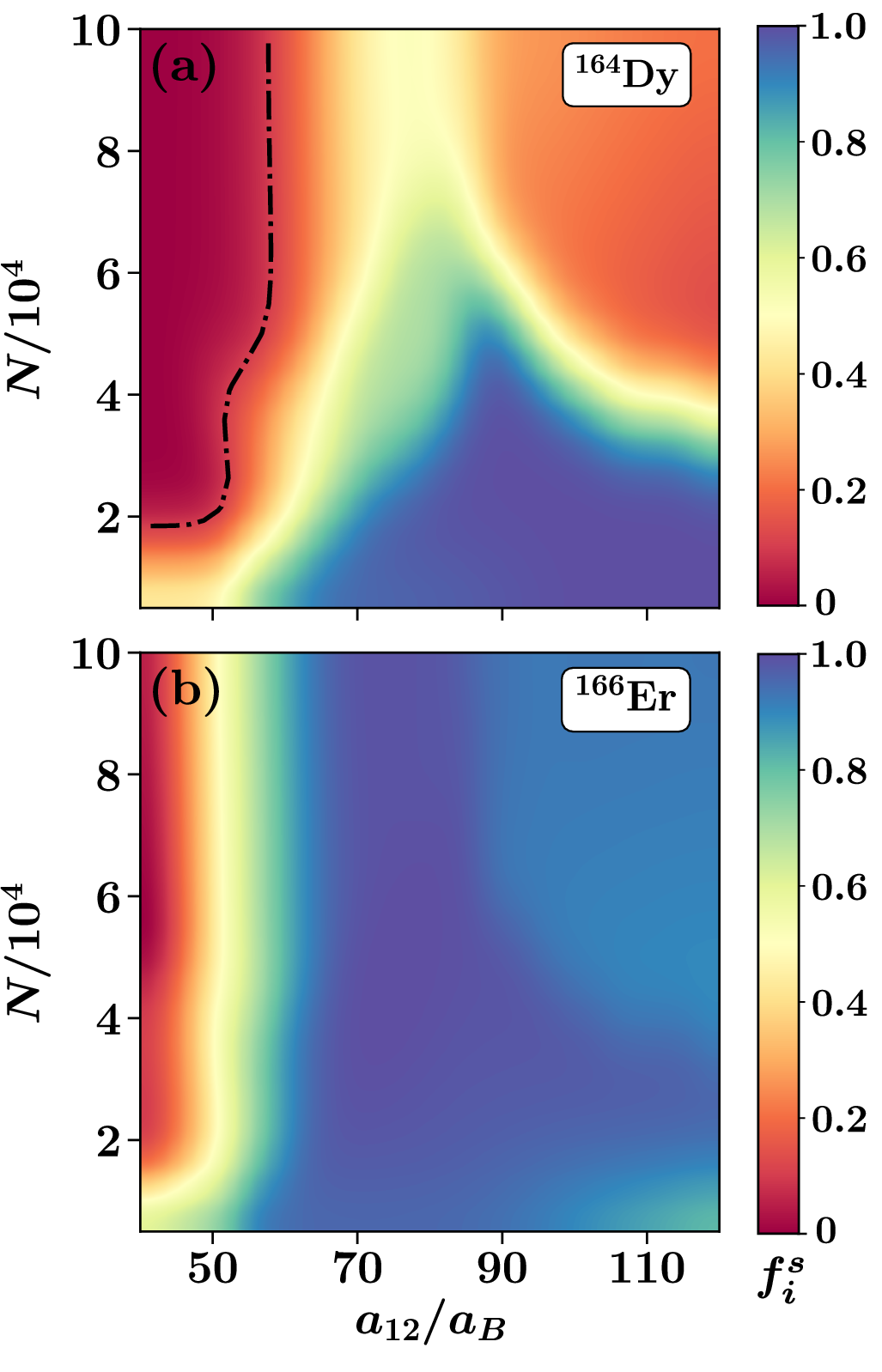}
	\caption{Superfluid fraction $f_i^s$ of (a) $^{164}$Dy $(f_1^s)$ and (b) $^{166}$Er condensates $(f_2^s)$ as a function of inter-species scattering length $a_{12}$ and number of particles in each condensate $N$. The black dash-dotted contour is drawn at $f_1^s=0.1$. The region inside the contour represents ID phase domain. No such region is present for Er condensate. Other parameters are same as Fig. \ref{fig:2}.}\label{fig:3}
\end{figure}
Nevertheless, the mere presence of a crystal-like structure alone does not conclusively indicate a SS phase. To better understand the degree of supersolidity, the superfluid fraction serves as a valuable measure. In the $x-y$ plane, the superfluid fraction is a rank-2 tensor with $x$ and $y$ as the eigen axes. The superfluid fraction of species-$i$ along the direction $\sigma$ is measured by applying a perturbation $-v_i^{\sigma} \hat{P}_i^{\sigma}$ to the species-$i$, where $v_i^{\sigma}$ and $\hat{P}_i^{\sigma}$ are the velocity and momentum operator, respectively\footnote[3]{This is equivalent to solving the stationary state in a comoving reference frame moving with a velocity $v_i^{\sigma}$ along the axis $\sigma$, i.e., $i\hbar\dot{\psi_i}=\big(\mathcal{H}_i^{GP}-v_i^{\sigma} \hat{P}_i^{\sigma}\big)\psi_i$.}. The superfluid fraction of the species-$i$ along the direction $\sigma$ is given by
\begin{equation}
	f_i^{s,\sigma}=1-\lim_{v_i^{\sigma}\to 0}\frac{\expval{\hat{P}_i^{\sigma}}}{N_i mv_i^{\sigma}},
\end{equation}
where $\expval{\hat{P}_i^{\sigma}}=-i\hbar\int\psi_i^{*}\partial \psi_i/\partial \sigma$ is the expectation value of momentum and $N_imv_i^{\sigma}$ is the total momemtum of species-$i$.
In our analysis, we employ Leggett's upper bound within the central region of each condensate of size $2L\times 2L$ to estimate the superfluid fraction along the axis $\sigma$ \cite{leggett_1970_can, leggett_1998_superfluid, chauveau_2023_superfluid}
\begin{align}
    f_i^{s,\sigma}&=\Bigg[\expval{n_i(\sigma)}\expval{\frac{1}{n_i(\sigma)}}\Bigg]^{-1}\nonumber\\&=(2L)^2\Bigg[\int_{-L}^{L}\dd\sigma~ n_i(\sigma)\int_{-L}^{L}\dd\sigma \frac{1}{n_i(\sigma)}\Bigg]^{-1},\label{f_s}
\end{align}
where $\sigma=\{x,y\}$, $x,y\in [-L,L]$ and $n_i(x)=\int\int\dd y \dd z \abs{\psi_i(x,y,z)}^2$. In our case, we consider $L=4.5\mu\rm m$. The overall superfluid fraction of each component is estimated by taking an average of the superfluid fraction along $x$ and $y$ direction $f_i^s=(f_i^{s,x}+f_i^{s,y})/2$. In this work, we categorize a component as an ID when its average superfluid fraction $f_i^s$ is less than 0.1. Conversely, a component with $f_i^s>0.1$, accompanied by a periodic density modulation, is considered to be in a SS state. Additionally, a component with $f_i^s>0.1$ and an unmodulated density profile is classified as a SF state. The superfluid fraction of each component is shown in Fig. \ref{fig:3}. 
%%%%%%%%%%%%%%%%%%%%%%%%%%%%%%%%%%%%%%%%%%%%%%%%%
\begin{figure}[tb!]
	\centering
	\includegraphics[width=0.45\textwidth]{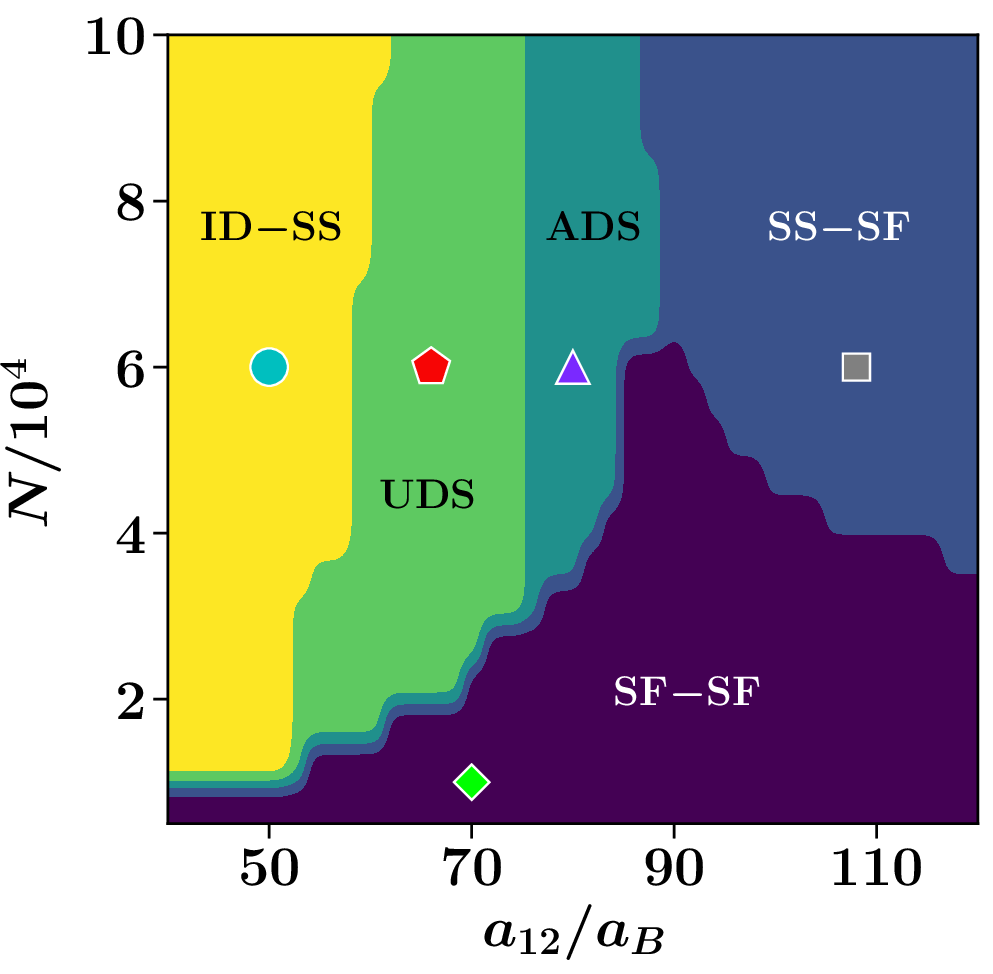}
	\caption{Ground state phase diagram of Dy-Er mixture in an oblate trap as a function of inter-species scattering length $a_{12}$ and number of particles in each condensate $N$. Different color domains highlighted with different markers represent SF-SF (\protect\marker{lightgreen}{diamond}), ID-SS (\protect\marker{cyan}{circle}), uniform domain supersolid (UDS \protect\markerpentagon{red}), alternating domain supersolid (ADS \protect\markertriangle{lightviolet}), and SS-SF (\protect\markersquare{gray}{rectangle}) phase domains, respectively. The result is for the case of binary dipolar BEC consisting of $^{164}$Dy and $^{166}$Er atoms with $a_{11}=90a_B$, $a_{11}^{\rm dd}=131a_B$ $(\mu_1^m=9.93\mu_B)$ and $a_{22}=80a_B$, $a_{22}^{\rm dd}=65.5a_B$ $(\mu_2^m=7\mu_B)$, respectively. The binary mixture is confined in a harmonic trap with $(\omega_x,\omega_y,\omega_z)=2\pi\times(45,45,133)$ Hz.}\label{fig:4}
\end{figure}
\subsection{Phase diagram}\label{seciiib}
\begin{figure}[H] 
	\centering
	\includegraphics[width=\textwidth]{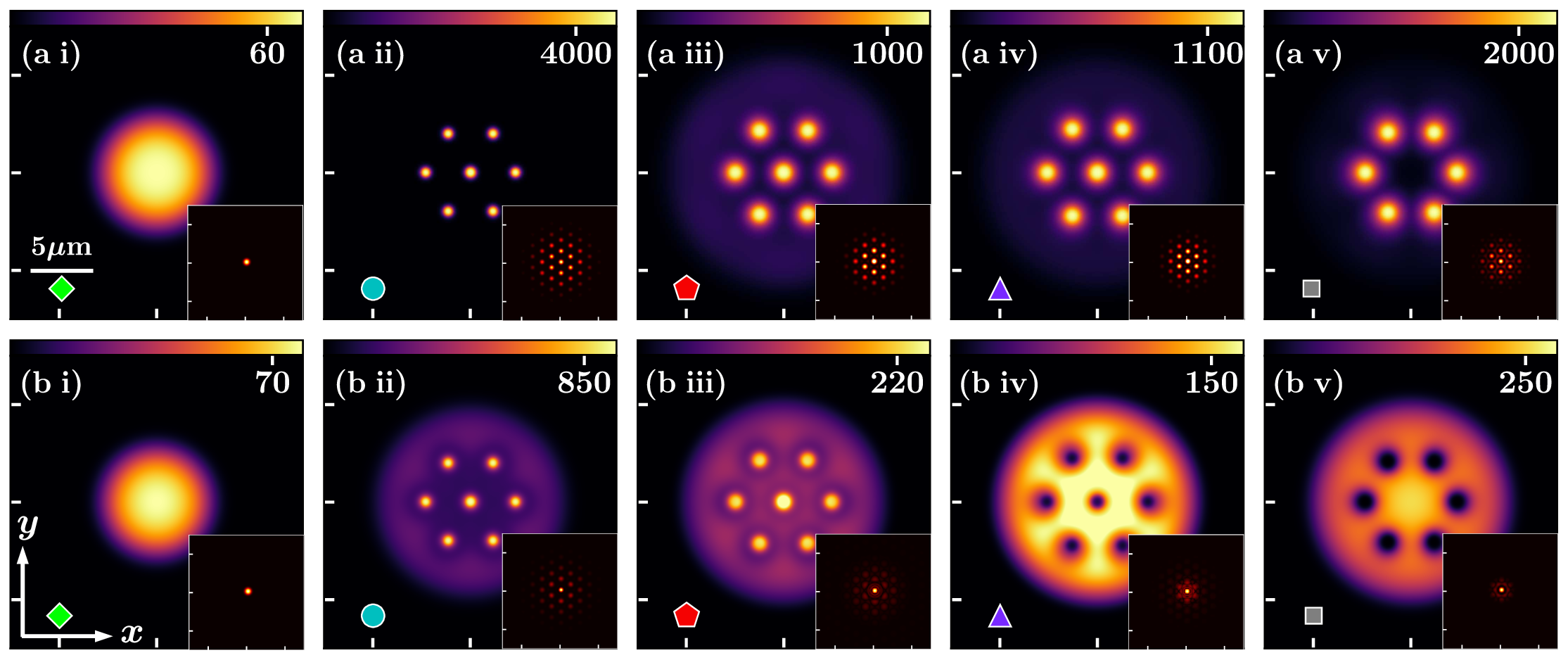}
		\caption{Ground state density profile of Dy-Er mixture. The upper and lower panel represent the density distributions of $^{164}$Dy and $^{166}$Er condensate in the $x-y$ plane, respectively. Each column represents a density distribution corresponding to (a i, b i) SF-SF (\protect\marker{lightgreen}{diamond}), (a ii, b ii) ID-SS (\protect\marker{cyan}{circle}), (a iii, b iii) UDS (\protect\markerpentagon{red}), (a iv, b iv) ADS (\protect\markertriangle{lightviolet}) and (a v, b v) SS-SF (\protect\markersquare{gray}{rectangle}) ground state phases, respectively. Locations of all these phases are highlighted in the ground state phase diagram by the corresponding marker (see Fig. \ref{fig:4}). The color bar denotes the density in units of $\mu\rm m^{-3}$. The insets in the lower right corners of each density distribution represent the power spectrum $S_i(k_x,k_y)$ in the $k_x-k_y$ plane.}\label{fig:5}
\end{figure}

In Fig. \ref{fig:4}, we depict the ground state phase diagram of the $^{164}$Dy-$^{166}$Er mixture. For a small number of particles, the binary mixture forms a SF-SF mixture [Fig. \ref{fig:4} \protect\marker{lightgreen}{diamond}] characterized by a smooth unmodulated density distribution on the $x-y$ plane [see Figs. \ref{fig:5}(a i) and \ref{fig:5}(b i)]. As we increase the number of particles, the anisotropic DDI strongly comes into play. At low $a_{12}$ (in the miscible phase domain $\Lambda>0.8$), due to the large intra-species anisotropic DDI, the atoms of $^{164}$Dy condensate experience an attractive potential at certain positions which leads to an increase in the density at those locations and consequently form a ID state [see Fig. \ref{fig:5}(a ii)] with a superfluid fraction $f^s_1<0.1$. As a consequence of these isolated density humps in the $^{164}$Dy condensate, the atoms of $^{166}$Er condensate occupying those positions are also experiencing significant attractive inter-species DDI. This results in the formation of periodic density humps (droplets) within the $^{166}$Er condensate. These induced droplets are coupled with each other via a background superfluid characterized by a lower peak density and higher superfluid fraction $f_2^s>0.1$ [see Fig. \ref{fig:5}(b ii)]. The binary mixture collectively forms a miscible ID-SS mixed state within this phase domain [Fig. \ref{fig:4} \protect\marker{cyan}{circle}].\par 

Increasing $a_{12}$ up to the miscible to immiscible phase transition boundary induces a transition to the uniform domain supersolid (UDS) state [Fig. \ref{fig:4} \protect\markerpentagon{red}]. Within this phase regime, the first species composed of $^{164}$Dy atoms forms a SS state with $f^s_1>0.1$ [Fig. \ref{fig:5} a iii]. Similar to the previous scenario, the $^{164}$Dy condensate induces supersolidity in the $^{166}$Er condensate as well. Notably, both condensates exhibit droplets located precisely at the same positions and these droplets are interconnected by a background superfluid [see Figs. \ref{fig:5}(a iii) and \ref{fig:5}(b iii)]. Nevertheless, the peak density of both condensates is lower compared to the peak density in the SS state of a single-component dipolar BEC $(a_{12}=0)$. Specifically, the peak density of the $^{166}$Er condensate is approximately an order of magnitude smaller than that of the $^{164}$Dy condensate and has an enhanced superfluid fraction $(f^s_2>0.7)$. Though we have not considered the three-body recombination loss, the reduced peak density of this UDS state leads to a decrease in the three-body recombination loss rate. As a consequence, the UDS state will persist for a longer time.\par 

As the value of $a_{12}$ is further increased, the overlap between the binary components diminishes. We observe a phase domain in close proximity to the miscible to immiscible phase boundary, where the relatively large inter-species repulsive interaction leads to the emergence of contrasting density patterns between the binary components. Specifically, while one component exhibits a density maximum, the other component forms a density minimum. This results in the formation of an intriguing alternating domain of supersolid (ADS) state [Fig. \ref{fig:4} \protect\markertriangle{lightviolet}]. In this phase regime, the $^{164}$Dy condensate adopts a hexagonal-shaped supersolid state [see Fig. \ref{fig:5}(a iv)] with a superfluid fraction $f^s_1>0.2$, whereas the $^{166}$Er condensate forms a honeycomb supersolid phase [see Fig. \ref{fig:5}(b iv)]. Remarkably, in this phase domain, the $^{166}$Er condensate exhibits a lower peak density resembling that of a SF state and also displays an enhanced superfluid fraction $(f^s_2>0.8)$.\par

Further increasing $a_{12}$ beyond the ADS phase domain, the inter-species DDI becomes repulsive. As a consequence, the $^{166}$Er condensate adopts a SF state, while the $^{164}$Dy condensate exhibits a SS state due to the attractive intra-species DDI. The $^{166}$Er condensate, owing to its relatively small intra-species $s$-wave scattering length, occupies the central position within the trap, while the $^{164}$Dy condensate forms a ring-shaped supersolid configuration. The ring supersolid state exhibits a lower superfluid fraction ($f^s_1>0.1$) compared    
{\unskip\parfillskip 0pt\par}

\onecolumngrid

\twocolumngrid
\noindent to the SS states observed in the UDS and ADS phases. The density distributions of the $^{164}$Dy and $^{166}$Er condensate in this SS-SF phase domain [Fig. \ref{fig:4} \protect\markersquare{gray}{rectangle}] are shown in Figures \ref{fig:5}(a v) and \ref{fig:5}(b v), respectively. It is worth noting that the occurrence of SS-SF and ADS states has also been recently predicted for a dipolar-nondipolar mixture and an anti-aligned dipolar mixture \cite{li_2022_longlifetime, bland_2022_alternatingdomain}.\par 
To further validate the crystalline structure we compute the power spectrum of each species $S_i(k_x,k_y)=\abs{\mathcal{F}[n_i(x,y,z=0)](k_x,k_y)}^2$, where $n_i(x,y,z=0)$ is the density of species-$i$ in the $z=0$ plane and $\mathcal{F}[n_i(x,y,0)](k_x,k_y)$ represent the Fourier transformation of the corresponding density distribution from position space to the momentum space. There is no additional density modulation along the axial direction, so examining the power spectrum in the $k_x-k_y$ plane is sufficient to comprehend the condensate's crystallinity. The insets of Fig. \ref{fig:5} represent the power spectrum in the $k_x-k_y$ plane. In the SF-SF mixture, a solitary peak is evident at $k_x=k_y=0$. However, in other states besides the central peak, we observe growths in the weight of the power spectrum at finite momentum values ($k_x,k_y\neq0$). These localized maxima in the power spectrum are organized in a triangular pattern, except for the $^{166}$Er condensate in the ADS and SS-SF mixed states. The triangular lattice pattern in the power spectrum signifies the emergence of multiple Brillouin zones formed by the droplets observed in the position space \cite{hertkorn_2021_pattern}. The power spectrum of the $^{166}$Er condensate in the ADS state exhibits a hexagonal pattern, visible at lower momenta, and is accompanied by a triangular lattice pattern at higher momentum values. In the SS-SF state, apart from the central maximum, the $^{166}$Er condensate showcases a hexagonal-shaped power spectrum at lower momentum values, attributed to the arrangement of density voids. The absence of a triangular lattice pattern indicates a lack of crystalline properties, signifying that the $^{166}$Er condensate is in the SF state.

\subsection{Interaction energies}\label{seciiic}
The interplay between intra- and inter-species CI and DDI results in the emergence of the above distinct ground state phases. The total energy of a Dy-Er mixture can be written as follows:
\begin{align}
    E=&\int \dd{\vb{r}} \Bigg[\sum_{i=1}^2\Bigg\{\frac{\hbar^2}{2m_i}\abs{\grad \psi_i(\vb{r})}^2 +V_t\abs{\psi_i(\vb{r})}^2\nonumber\Bigg\}\Bigg]\\&+\sum_{i,j=1}^2\Big(E^{\rm{CI}}_{ij}+E^{\rm{DDI}}_{ij}\Big)+\frac{2}{5}\gamma_1(\epsilon_{11}^{\rm{dd}})\int\dd{\vb{r}}\abs{\psi_1(\vb{r})}^5.\label{energy}
\end{align}
Here the first two terms represent the kinetic energy and external potential energy due to the harmonic confinement, respectively. The last term represents the beyond mean-field LHY interaction energy for the Dy condensate\footnote[5]{Due to the low peak density of the $^{166}$Er condensate, it does not require quantum stabilization. However, when taking into account the LHY correction for the $^{166}$Er condensate, the energy contribution attributed to the LHY correction is negligible. Therefore, we did not include the energy contribution due to the LHY correction for the $^{166}$Er condensate in the total energy expression.}. The energy terms $E^{\rm CI}_{ij}$ and $E^{\rm DDI}_{ij}$ refer to CI and DDI energy, respectively and can be written as
\begin{align}
    E^{\rm {CI}}_{ij}= \frac{1}{2}\int \dd{\vb{r}} g_{ij}\abs{\psi_i(\vb{r})}^2 \abs{\psi_j(\vb{r})}^2,
\end{align}
and
\begin{align}
    E^{\rm{DDI}}_{ij}=\frac{1}{2}\int\int \dd{\vb{r}}\dd{\vb{r}^{\prime}} V_{ij}^{\rm {dd}}(\vb{r}-\vb{r^{\prime}})\abs{\psi_j(\vb{r^{\prime}})}^2\abs{\psi_i(\vb{r})}^2.
\end{align}
\begin{figure}[tb!]
	\centering
	\includegraphics[width=0.46\textwidth]{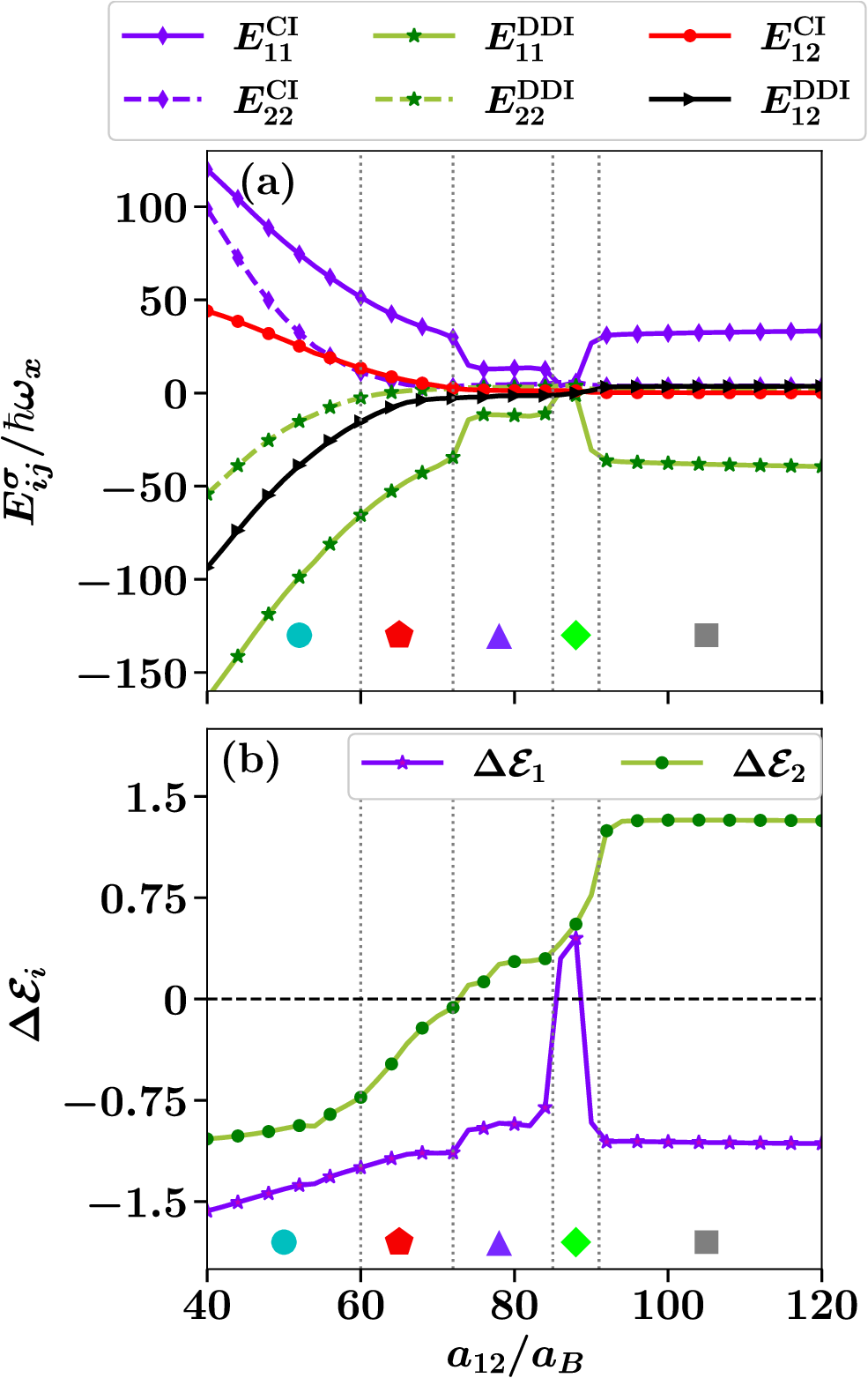}
		\caption{(a) Variation of different intra- and inter-species energy components $E^{\sigma}_{ij}$ in units of $\hbar\omega_x$ with the inter-species scattering length $a_{12}$. $\sigma$ denotes CI and DDI (see the legend). Panel (b) shows the variation of $\Delta\mathcal{E}_i$ with $a_{12}$. $\Delta\mathcal{E}_i$ quantifies the relative strength of DDI energy over the CI energy. The dotted vertical lines separate different phase domains and are marked by the corresponding markers. Results are for the case of $N=6\times 10^4$ number of particles in each condensate. Other parameters are same as of Fig. \ref{fig:4}. }\label{fig:6}
\end{figure}
 Figure \ref{fig:6}(a) illustrates the variation of intra-species and inter-species CI energies as well as DDI energies with $a_{12}$ for $N=6\times10^4$ number of particles in each condensate. To quantify the relative strength of DDI energy over the CI energy for the species-$i$, we introduce a dimensionless parameter $\Delta\mathcal{E}_i=\sum_j E^{\rm{DDI}}_{ij}/\sum_j E^{\rm{CI}}_{ij}$. In Fig. \ref{fig:6}(b), we have shown the variation of $\Delta\mathcal{E}_i$ with the inter-species scattering length $a_{12}$. In general, the value of $\Delta\mathcal{E}_i<-1$ implies a dominating attractive DDI energy characterizing a self-bound ID phase. Conversely, $\Delta\mathcal{E}_i>0$ signifies a SF phase with dominating repulsive interactions. In the SS state, the dimensionless parameter $\Delta\mathcal{E}_i$ attains an intermediate value between $-1$ to $0$. Additionally, we observe that a honeycomb supersolid state appears when $\Delta\mathcal{E}_i$ slightly trends toward the positive value (in the vicinity of $\Delta\mathcal{E}_i\to 0$).\par

For very small values of $a_{12}$, the DDI energy dominates over the CI energy, leading to $\Delta\mathcal{E}_1<-1$ for the $^{164}$Dy condensate, while $\Delta\mathcal{E}_2\geq-1$ for the $^{166}$Er condensate due to smaller intra-species DDI energy. Consequently, the $^{164}$Dy condensate adopts an ID state and the $^{166}$Er condensate forms a SS state. The value of $\Delta\mathcal{E}_i$ for both the condensates increases with the increase of $a_{12}$. Within the UDS phase domain, both condensates exhibit $-1.1\leq\Delta\mathcal{E}_i\leq0$ and form a binary SS state. In the ADS phase domain, although the value of $\Delta\mathcal{E}_1<0$, the $\Delta\mathcal{E}_2$ becomes slightly positive. As a result of this, the superfluidity of the $^{166}$Er condensate increases and adopts a hexagonal supersolid state. There exists a narrow window where both components have positive $\Delta\mathcal{E}_i$ and form a SF-SF mixture. Further increasing $a_{12}$, the atoms of $^{164}$Dy condensate experience an attractive intra-species DDI energy $(\Delta\mathcal{E}_1\leq-1)$, whereas the $^{166}$Er condensate experience a dominating repulsive interaction energy $(\Delta\mathcal{E}_2>1)$. This results in the formation of a SS-SF mixed state.

%%%%%%%%%%%%%%%%%%%%%%%%%%%%%%%%%%%%%%%%%%%%
%%%%%% DYNAMICS %%%%%%%%

\section{Quench induced dynamics of \texorpdfstring{D\MakeLowercase{y}-E\MakeLowercase{r}}{Dy-Er} mixture}\label{seciv}
So far, our discussion has centered around the potential phases that the $^{164}$Dy-$^{166}$Er mixture could adopt in its ground state. Our focus will now shift toward investigating the dynamical phase transition across these phase boundaries. This transition is triggered by the quenching of the inter-species scattering length $a_{12}$.\par
\begin{figure}[tb!]
	\centering
	\includegraphics[width=0.48\textwidth]{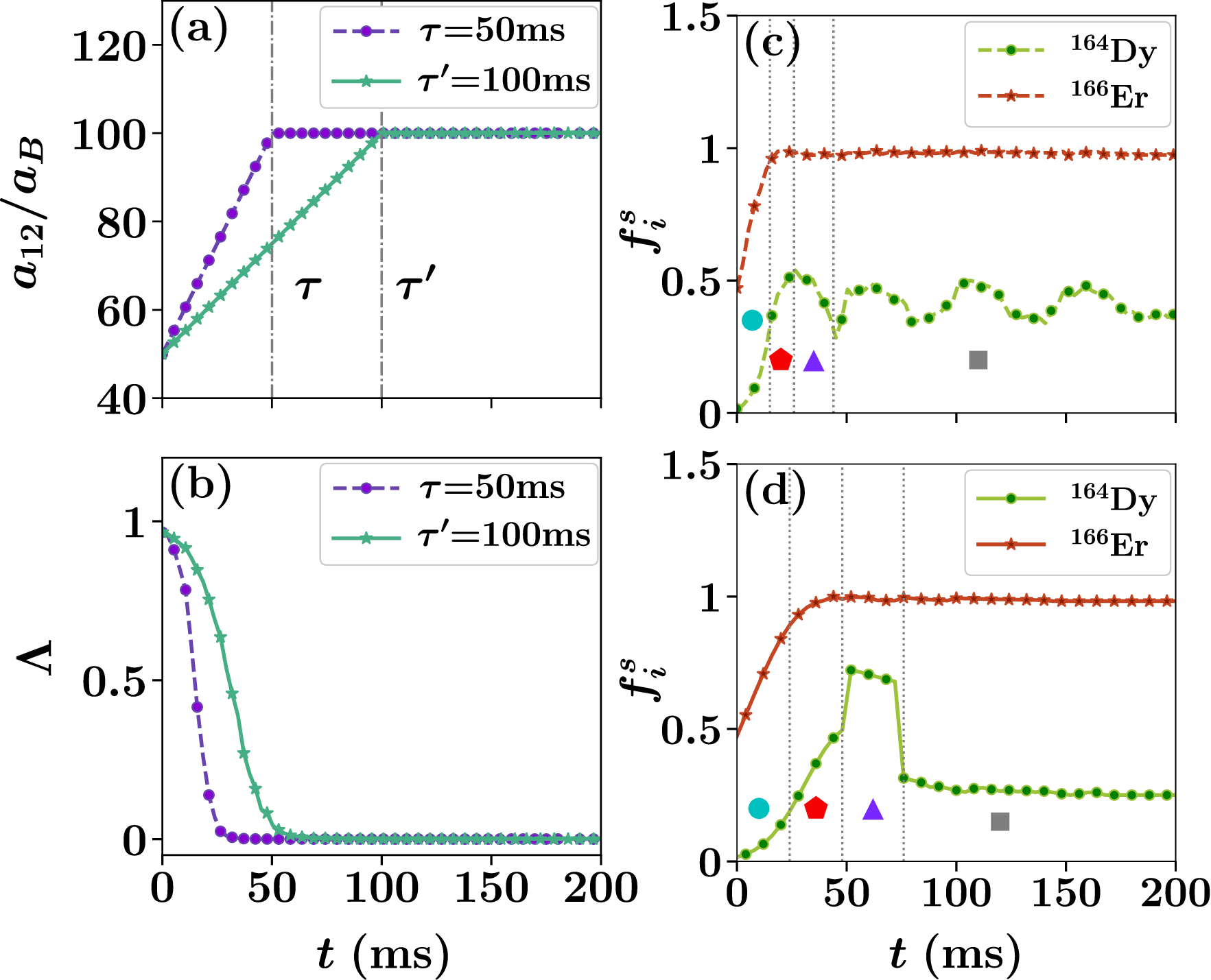}
		\caption{Quenching of inter-species scattering length $a_{12}$ from an initial value $50a_B$ to a final value $100a_B$ of a $^{164}$Dy-$^{166}$Er mixture consist of $N_1=N_2=6\times10^4$ number of particles in each condensate. Panel (a) shows two different linear ramps with ramp time $\tau=50\rm ms$ and $\tau^{\prime}=100\rm ms$. Panel (b) shows the corresponding variation of the value of the overlap integral $\Lambda$. Panel (c) and (d) show the time evolution of superfluid fraction $f_i^s$ of each condensate due to the quenching of $a_{12}$ in $\tau=50\rm ms$ and $\tau^{\prime}=100\rm ms$, respectively. The dotted vertical lines in panels (c) and (d) separate different phases that emerge due to the quench-induced dynamics and are marked by the corresponding marker. Other parameters are same as of Fig. \ref{fig:4}.}\label{fig:7}
\end{figure}

 
To initiate this study, we first prepare the $^{164}$Dy-$^{166}$Er mixture in a miscible ID-SS configuration, with interaction parameters set as follows: $a_{11}=90a_B$, $a_{22}=80a_B$, and $a_{12}=50a_B$. Each condensate contains a total of $N=6\times10^4$ particles. A small-amplitude noise is introduced to the initial state to incorporate the influence of thermal fluctuations. Subsequently, we gradually increase the inter-species scattering length from its initial value of $a_{12}=50a_B$ to a higher value of $a_{12}=100a_B$. This modulation of $a_{12}$ is executed using two distinct linear ramps with ramp time $\tau=50\rm{ms}$ and $\tau^{\prime}=100\rm{ms}$ and after which $a_{12}$ is kept constant to check the stability of the evolved state [see Fig. \ref{fig:7}(a)].\par 

In Figure \ref{fig:7}(b) and \ref{fig:7}(c,d), we present the temporal evolution of the overlap integral $\Lambda$ and the superfluid fraction $f_i^{s}$ for each species, corresponding to both the ramps with ramp times $\tau$ and $\tau^{\prime}$, respectively. Initially, owing to the relatively small inter-species interactions, the binary mixture adopts a miscible configuration where $\Lambda$ attains unity. During this phase, the $^{164}$Dy condensate maintains a superfluid fraction $f^s_1<0.1$, while the $^{166}$Er condensate exhibits a superfluid fraction around $f^s_2\approx0.5$. Throughout both the quenching process, the overlap integral experiences a gradual reduction while the superfluid fraction of both condensates increases [see Fig. \ref{fig:7}(b)]. Remarkably, the $^{166}$Er condensate consistently maintains a higher superfluid fraction than the $^{164}$Dy condensate as the dynamics unfold. It is also interesting to note that the superfluid fraction of the $^{166}$Er condensate exhibits a smooth transition across all phase boundaries [see Fig. \ref{fig:7}(c, d)]. In contrast, the superfluid fraction of the $^{164}$Dy condensate follows a distinct pattern: it initially increases up to the phase boundary separating the UDS and ADS states and then subsequently decreases up to the boundary between ADS and SS-SF phases [see Fig. \ref{fig:7} (c, d)]. In between the transitions from UDS to ADS and ADS to SS-SF phases, we find an abrupt alteration in the value of the superfluid fraction of the $^{164}$Dy condensate [Fig. \ref{fig:7} (d)]. However, these abrupt shifts are absent when quenching is done with quicker ramp time $(\tau=50\rm ms)$ instead, the superfluid fraction of the $^{164}$Dy condensate oscillates beyond the ADS to SS-SF phase transition boundary [Fig. \ref{fig:7} (c)].\par
\begin{figure}[tb!]
	\centering
	\includegraphics[width=0.48\textwidth]{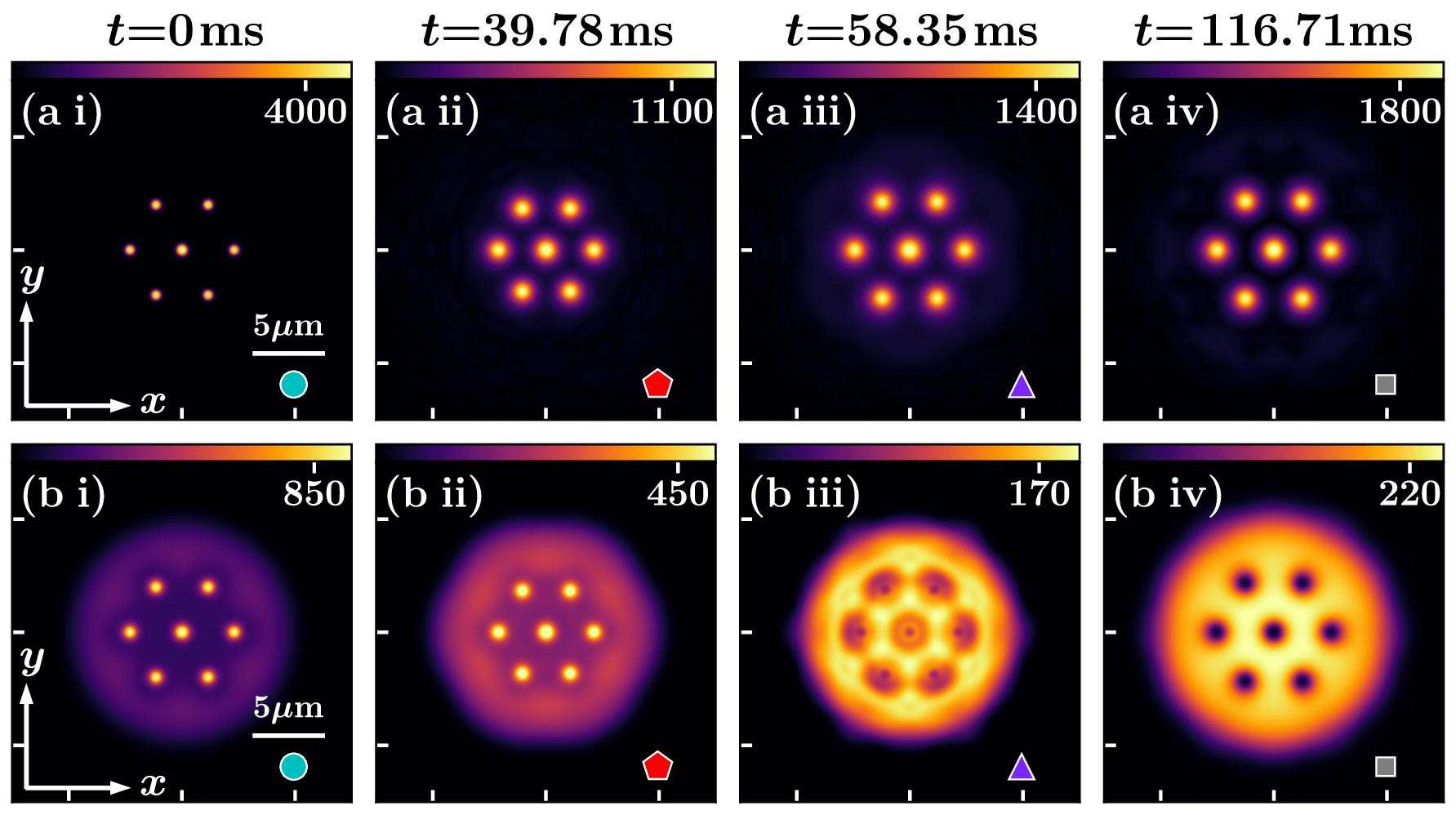}
		\caption{Snapshots of the density distributions of Dy-Er mixture following the quenched-induced dynamics with ramp time $\tau^{\prime}=100\rm ms$ in the $x-y$ plane. The upper and the lower panels represent the density $n(x,y,z=0,t)$ of $^{164}$Dy and $^{166}$Er condensate, respectively. Each column represents a different phase and is marked by the corresponding marker to denote (a i, b i) ID-SS (\protect\marker{cyan}{circle}), (a ii, b ii) UDS (\protect\markerpentagon{red}), (a iii, b iii) ADS (\protect\markertriangle{lightviolet}) and (a iv, b iv) SS-SF (\protect\markersquare{gray}{rectangle}) states. The color bar denotes the density in units of $\mu\rm m^{-3}$.}\label{fig:8}
\end{figure}
The SS-SF state produced through these quench-induced dynamics [see Fig. \ref{fig:8}(a iv), (b iv)] is slightly different from the SS-SF state observed in the ground state. In the ground state, the $^{164}$Dy condensate adopts a ring-like supersolid configuration. Whereas, following the quench-induced dynamics the $^{164}$Dy condensate forms a metastable state due to the transition through discontinuous phase boundaries. It exhibits six droplets interconnected by a background superfluid, forming a ring-like arrangement with an isolated droplet positioned at the central point of the ring [see Fig. \ref{fig:8}(a iv)]. Meanwhile, due to dominating repulsive interactions, the $^{166}$Er condensate adopts a superfluid state with certain density voids following both quenching processes [see Fig. \ref{fig:8}(b iv)]. The representative instantaneous density profiles of different states produced during the interaction quench ($\tau=100\rm ms$) are shown in Fig. \ref{fig:8}(a i - b iv).

% %%%%%%%%%%%%%%%%%%%%%%%%%%%%%%%%%%%%%%%%%%%%%%
% %Conclusions

\section{Conclusions and outlook}\label{secv}
In conclusion, we demonstrate that though the $^{166}$Er condensate does not primarily exhibit a dipole-dominated interaction within a $^{164}$Dy-$^{166}$Er mixture, the presence of the $^{164}$Dy condensate can trigger the formation of a quantum droplet and induce supersolidity in $^{166}$Er condensate. The imbalance in the dipole-dipole interaction strength, the interplay between the various intra-species and inter-species interactions and the anisotropic coupling between the components results in a wide range of unique groundstate phases. We have reported the emergence of fascinating binary SS configurations, encompassing uniform and alternating domain supersolid (UDS and ADS) states and mixed states that include combinations of ID-SS and SS-SF states. We characterize all these phases and construct a ground state phase diagram for a wide range of parameters within the current experimental reach. In contrast to single-component dipolar and dipolar-nondipolar mixtures, the heteronuclear mixture of $^{164}$Dy and $^{166}$Er condensates displays a broader parameter regime where various SS states emerge. In these SS phases, both condensates maintain comparatively low peak densities, potentially leading to prolonged lifetimes of these phases. Notably, owing to a relatively weak intra-species DDI strength and a dominant repulsive intra-species contact interaction, the $^{166}$Er condensate results in nearly an order of magnitude lower peak density than the $^{164}$Dy condensate. In fact, the induced SS states in $^{166}$Er condensate exhibit a peak density resembling that of a superfluid state. Consequently, to form these unique SS states, the $^{166}$Er condensate does not require quantum stabilization in the form of LHY correction. Owing to the diverse interactions, including intra- and inter-species interactions (CI and DDI), the heteronuclear binary mixture is also capable of exotic pattern formation. In addition to the trivial hexagonal supersolid structure with a triangular droplet lattice, $^{166}$Er condensate possesses a unique honeycomb supersolid pattern within the alternating domain supersolid phase. Furthermore, the $^{164}$Dy condensate adopts a ring-like supersolid structure in a SS-SF mixed phase. We also monitor the dynamical nucleation of these phases, starting from an initial ID-SS mixed state to a SS-SF mixed state. This transition takes place through the UDS and ADS intermediate states and is achieved via linear quenches of the inter-species scattering length. Overall, our findings demonstrate that the heteronuclear binary dipolar mixture offers an intriguing platform for observing a variety of unique long-lived supersolid states and diverse physical phenomena within the current experimental reach. \par

Our observations pave the way for numerous promising research directions for future endeavors. One straightforward option is to investigate the lifetime of these unique doubly SS (UDS and ADS) and mixed phases (ID-SS and SS-SF) by incorporating the three-body interaction loss \cite{wachtler_2016_quantum}. Another important aspect will be to study the elementary excitations \cite{wachtler_2016_groundstate, diniz_2020_ground} of these phases and across the phase transition boundaries. In our work, we have restricted our investigation to a $^{164}$Dy-$^{166}$Er mixture. However, Dy and Er feature many stable bosonic and fermionic isotopes \cite{trautmann_2018_dipolar} and it would be intriguing to explore possible physical phenomena in other Bose-Bose, Bose-Fermi and Fermi-Fermi mixtures \cite{ferrier-barbut_2014_mixture, ruiz-tijerina_2023_bose, ravensbergen_2020_resonantly}. Another vital prospect would be to investigate the exotic pattern formation \cite{hertkorn_2021_pattern, zhang_2021_phases, saboo_2023_rayleightaylor} and various topological excitations such as the formation of solitons and vortex clusters \cite{raghunandan_2015_twodimensional, klaus_2022_observationa, das_2022_vortex} in a heteronuclear binary dipolar mixture. Moreover, the direct dynamical nucleation of these phases via the evaporation cooling process and investigation of the impact of finite temperature on these phases would be highly interesting \cite{bland_2022_twodimensional, sanchez-baena_2023_heating}.
%%%%%%%%%%%%%%%%%%%%%%%%%%%%%%%%%%%%%%%%%%%%%%%%%%%%%%%%%%%%
%Acknowledgements

\begin{acknowledgments}
 We acknowledge the National Supercomputing Mission (NSM) for providing computing resources of ‘PARAM Shakti’ at IIT Kharagpur, which is implemented by C-DAC and supported by the Ministry of Electronics and Information Technology (MeitY) and Department of Science and Technology (DST), Government of India. S. Halder acknowledges the MHRD Govt. of India for the research fellowship. S. Das acknowledges support from AFOSR FA9550-23-1-0034.
\end{acknowledgments}

% %%%%%%%%%%%%%%%%%%%%%%%%%%%%%%%%%%%%%%
\appendix
\section{Numerical Methods}\label{A}
Results in this work are based on three-dimensional numerical simulations in the coupled eGPE framework. For the sake of the convenience of numerical simulations and better computational precision, we cast the coupled eGPE into a dimensionless form. This is achieved by rescaling the length scale and time scale in terms of oscillator length $l_{osc}=\sqrt{\hbar/m\omega_x}$ and $\omega_x$ trapping frequency along the x direction. Under this transformation, the wave function of species-$i$ obeys $\psi_i(\vb{r}^{\prime})=\sqrt{l_{osc}^3/N_i}\psi_i(\vb{r})$, where $N_i$ is the number of particles in species-$i$. After the transformation of variables into dimensionless quantities, the coupled eGPE is solved by split-step-Crank-Nicolson scheme \cite{crank_1947_practical}. Since the dipolar potential has a singularity at $r=0$ (see Eq. (\ref{dip_pot})), it is numerically evaluated in Fourier space and we obtain the real space contribution through the application of the convolution theorem. The groundstates of binary dipolar condensate are obtained by propagating the relevant equations in imaginary time until the relative deviations of the wave functions (calculated at every grid point) and energy of each condensate between successive time steps are less than $10^{-6}$ and $10^{-7}$, respectively. Furthermore, we fix the normalization of each species at every time instant of the imaginary time propagation. Using this groundstate solution as an initial state, at $t=0$, and by changing the interaction strengths, we monitor their evolution in real-time. Our simulations are performed within a 3D box grid containing $(256\times 256\times 128)$ grid points, with the spatial grid spacing $\Delta_x=\Delta_y=0.1 l_{osc}$ and $\Delta_z=0.2 l_{osc}$ while the time step $\Delta_t=2\times 10^{-4}/\omega_x$.
% %%%%%%%%%%%%%%%%%%%%%%%%%%%%%%%%%%%%%%%%%%%%%%%%%
\bibliographystyle{apsrev4-2}
\bibliography{reference} 

\begin{thebibliography}{100}%
\makeatletter
\providecommand \@ifxundefined [1]{%
 \@ifx{#1\undefined}
}%
\providecommand \@ifnum [1]{%
 \ifnum #1\expandafter \@firstoftwo
 \else \expandafter \@secondoftwo
 \fi
}%
\providecommand \@ifx [1]{%
 \ifx #1\expandafter \@firstoftwo
 \else \expandafter \@secondoftwo
 \fi
}%
\providecommand \natexlab [1]{#1}%
\providecommand \enquote  [1]{``#1''}%
\providecommand \bibnamefont  [1]{#1}%
\providecommand \bibfnamefont [1]{#1}%
\providecommand \citenamefont [1]{#1}%
\providecommand \href@noop [0]{\@secondoftwo}%
\providecommand \href [0]{\begingroup \@sanitize@url \@href}%
\providecommand \@href[1]{\@@startlink{#1}\@@href}%
\providecommand \@@href[1]{\endgroup#1\@@endlink}%
\providecommand \@sanitize@url [0]{\catcode `\\12\catcode `\$12\catcode
  `\&12\catcode `\#12\catcode `\^12\catcode `\_12\catcode `\%12\relax}%
\providecommand \@@startlink[1]{}%
\providecommand \@@endlink[0]{}%
\providecommand \url  [0]{\begingroup\@sanitize@url \@url }%
\providecommand \@url [1]{\endgroup\@href {#1}{\urlprefix }}%
\providecommand \urlprefix  [0]{URL }%
\providecommand \Eprint [0]{\href }%
\providecommand \doibase [0]{https://doi.org/}%
\providecommand \selectlanguage [0]{\@gobble}%
\providecommand \bibinfo  [0]{\@secondoftwo}%
\providecommand \bibfield  [0]{\@secondoftwo}%
\providecommand \translation [1]{[#1]}%
\providecommand \BibitemOpen [0]{}%
\providecommand \bibitemStop [0]{}%
\providecommand \bibitemNoStop [0]{.\EOS\space}%
\providecommand \EOS [0]{\spacefactor3000\relax}%
\providecommand \BibitemShut  [1]{\csname bibitem#1\endcsname}%
\let\auto@bib@innerbib\@empty
%</preamble>
\bibitem [{\citenamefont {Gross}(1957)}]{gross_1957_unified}%
  \BibitemOpen
  \bibfield  {author} {\bibinfo {author} {\bibfnamefont {E.~P.}\ \bibnamefont
  {Gross}},\ }\href {https://doi.org/10.1103/PhysRev.106.161} {\bibfield
  {journal} {\bibinfo  {journal} {Physical Review}\ }\textbf {\bibinfo {volume}
  {106}},\ \bibinfo {pages} {161} (\bibinfo {year} {1957})}\BibitemShut
  {NoStop}%
\bibitem [{\citenamefont {Leggett}(1970)}]{leggett_1970_can}%
  \BibitemOpen
  \bibfield  {author} {\bibinfo {author} {\bibfnamefont {A.~J.}\ \bibnamefont
  {Leggett}},\ }\href {https://doi.org/10.1103/PhysRevLett.25.1543} {\bibfield
  {journal} {\bibinfo  {journal} {Physical Review Letters}\ }\textbf {\bibinfo
  {volume} {25}},\ \bibinfo {pages} {1543} (\bibinfo {year}
  {1970})}\BibitemShut {NoStop}%
\bibitem [{\citenamefont {Toennies}\ \emph {et~al.}(2001)\citenamefont
  {Toennies}, \citenamefont {Vilesov},\ and\ \citenamefont
  {Whaley}}]{toennies_2001_superfluid}%
  \BibitemOpen
  \bibfield  {author} {\bibinfo {author} {\bibfnamefont {J.~P.}\ \bibnamefont
  {Toennies}}, \bibinfo {author} {\bibfnamefont {A.~F.}\ \bibnamefont
  {Vilesov}},\ and\ \bibinfo {author} {\bibfnamefont {K.~B.}\ \bibnamefont
  {Whaley}},\ }\href {https://doi.org/10.1063/1.1359707} {\bibfield  {journal}
  {\bibinfo  {journal} {Physics Today}\ }\textbf {\bibinfo {volume} {54}},\
  \bibinfo {pages} {31} (\bibinfo {year} {2001})}\BibitemShut {NoStop}%
\bibitem [{\citenamefont {Toennies}\ and\ \citenamefont
  {Vilesov}(2004)}]{toennies_2004_superfluid}%
  \BibitemOpen
  \bibfield  {author} {\bibinfo {author} {\bibfnamefont {J.~P.}\ \bibnamefont
  {Toennies}}\ and\ \bibinfo {author} {\bibfnamefont {A.~F.}\ \bibnamefont
  {Vilesov}},\ }\href {https://doi.org/10.1002/anie.200300611} {\bibfield
  {journal} {\bibinfo  {journal} {Angewandte Chemie International Edition}\
  }\textbf {\bibinfo {volume} {43}},\ \bibinfo {pages} {2622} (\bibinfo {year}
  {2004})}\BibitemShut {NoStop}%
\bibitem [{\citenamefont {Barranco}\ \emph {et~al.}(2006)\citenamefont
  {Barranco}, \citenamefont {Guardiola}, \citenamefont {Hern{\'a}ndez},
  \citenamefont {Mayol}, \citenamefont {Navarro},\ and\ \citenamefont
  {Pi}}]{barranco_2006_helium}%
  \BibitemOpen
  \bibfield  {author} {\bibinfo {author} {\bibfnamefont {M.}~\bibnamefont
  {Barranco}}, \bibinfo {author} {\bibfnamefont {R.}~\bibnamefont {Guardiola}},
  \bibinfo {author} {\bibfnamefont {S.}~\bibnamefont {Hern{\'a}ndez}}, \bibinfo
  {author} {\bibfnamefont {R.}~\bibnamefont {Mayol}}, \bibinfo {author}
  {\bibfnamefont {J.}~\bibnamefont {Navarro}},\ and\ \bibinfo {author}
  {\bibfnamefont {M.}~\bibnamefont {Pi}},\ }\href
  {https://doi.org/10.1007/s10909-005-9267-0} {\bibfield  {journal} {\bibinfo
  {journal} {Journal of Low Temperature Physics}\ }\textbf {\bibinfo {volume}
  {142}},\ \bibinfo {pages} {1} (\bibinfo {year} {2006})}\BibitemShut {NoStop}%
\bibitem [{\citenamefont {Boninsegni}\ and\ \citenamefont
  {Prokof'ev}(2012)}]{boninsegni_2012_colloquium}%
  \BibitemOpen
  \bibfield  {author} {\bibinfo {author} {\bibfnamefont {M.}~\bibnamefont
  {Boninsegni}}\ and\ \bibinfo {author} {\bibfnamefont {N.~V.}\ \bibnamefont
  {Prokof'ev}},\ }\href {https://doi.org/10.1103/RevModPhys.84.759} {\bibfield
  {journal} {\bibinfo  {journal} {Reviews of Modern Physics}\ }\textbf
  {\bibinfo {volume} {84}},\ \bibinfo {pages} {759} (\bibinfo {year}
  {2012})}\BibitemShut {NoStop}%
\bibitem [{\citenamefont {Chan}\ \emph {et~al.}(2013)\citenamefont {Chan},
  \citenamefont {Hallock},\ and\ \citenamefont {Reatto}}]{chan_2013_overview}%
  \BibitemOpen
  \bibfield  {author} {\bibinfo {author} {\bibfnamefont {M.~H.~W.}\
  \bibnamefont {Chan}}, \bibinfo {author} {\bibfnamefont {R.~B.}\ \bibnamefont
  {Hallock}},\ and\ \bibinfo {author} {\bibfnamefont {L.}~\bibnamefont
  {Reatto}},\ }\href {https://doi.org/10.1007/s10909-013-0882-x} {\bibfield
  {journal} {\bibinfo  {journal} {Journal of Low Temperature Physics}\ }\textbf
  {\bibinfo {volume} {172}},\ \bibinfo {pages} {317} (\bibinfo {year}
  {2013})}\BibitemShut {NoStop}%
\bibitem [{\citenamefont {Ancilotto}\ \emph {et~al.}(2017)\citenamefont
  {Ancilotto}, \citenamefont {Barranco}, \citenamefont {Coppens}, \citenamefont
  {Eloranta}, \citenamefont {Halberstadt}, \citenamefont {Hernando},
  \citenamefont {Mateo},\ and\ \citenamefont {Pi}}]{ancilotto_2017_density}%
  \BibitemOpen
  \bibfield  {author} {\bibinfo {author} {\bibfnamefont {F.}~\bibnamefont
  {Ancilotto}}, \bibinfo {author} {\bibfnamefont {M.}~\bibnamefont {Barranco}},
  \bibinfo {author} {\bibfnamefont {F.}~\bibnamefont {Coppens}}, \bibinfo
  {author} {\bibfnamefont {J.}~\bibnamefont {Eloranta}}, \bibinfo {author}
  {\bibfnamefont {N.}~\bibnamefont {Halberstadt}}, \bibinfo {author}
  {\bibfnamefont {A.}~\bibnamefont {Hernando}}, \bibinfo {author}
  {\bibfnamefont {D.}~\bibnamefont {Mateo}},\ and\ \bibinfo {author}
  {\bibfnamefont {M.}~\bibnamefont {Pi}},\ }\href
  {https://doi.org/10.1080/0144235X.2017.1351672} {\bibfield  {journal}
  {\bibinfo  {journal} {International Reviews in Physical Chemistry}\ }\textbf
  {\bibinfo {volume} {36}},\ \bibinfo {pages} {621} (\bibinfo {year}
  {2017})}\BibitemShut {NoStop}%
\bibitem [{\citenamefont {Kim}\ and\ \citenamefont
  {Chan}(2004)}]{kim_2004_probable}%
  \BibitemOpen
  \bibfield  {author} {\bibinfo {author} {\bibfnamefont {E.}~\bibnamefont
  {Kim}}\ and\ \bibinfo {author} {\bibfnamefont {M.~H.~W.}\ \bibnamefont
  {Chan}},\ }\href {https://doi.org/10.1038/nature02220} {\bibfield  {journal}
  {\bibinfo  {journal} {Nature}\ }\textbf {\bibinfo {volume} {427}},\ \bibinfo
  {pages} {225} (\bibinfo {year} {2004})}\BibitemShut {NoStop}%
\bibitem [{\citenamefont {Kim}\ and\ \citenamefont
  {Chan}(2012)}]{kim_2012_absence}%
  \BibitemOpen
  \bibfield  {author} {\bibinfo {author} {\bibfnamefont {D.~Y.}\ \bibnamefont
  {Kim}}\ and\ \bibinfo {author} {\bibfnamefont {M.~H.~W.}\ \bibnamefont
  {Chan}},\ }\href {https://doi.org/10.1103/PhysRevLett.109.155301} {\bibfield
  {journal} {\bibinfo  {journal} {Phys. Rev. Lett.}\ }\textbf {\bibinfo
  {volume} {109}},\ \bibinfo {pages} {155301} (\bibinfo {year}
  {2012})}\BibitemShut {NoStop}%
\bibitem [{\citenamefont {Cinti}\ \emph {et~al.}(2010)\citenamefont {Cinti},
  \citenamefont {Jain}, \citenamefont {Boninsegni}, \citenamefont {Micheli},
  \citenamefont {Zoller},\ and\ \citenamefont
  {Pupillo}}]{cinti_2010_supersolid}%
  \BibitemOpen
  \bibfield  {author} {\bibinfo {author} {\bibfnamefont {F.}~\bibnamefont
  {Cinti}}, \bibinfo {author} {\bibfnamefont {P.}~\bibnamefont {Jain}},
  \bibinfo {author} {\bibfnamefont {M.}~\bibnamefont {Boninsegni}}, \bibinfo
  {author} {\bibfnamefont {A.}~\bibnamefont {Micheli}}, \bibinfo {author}
  {\bibfnamefont {P.}~\bibnamefont {Zoller}},\ and\ \bibinfo {author}
  {\bibfnamefont {G.}~\bibnamefont {Pupillo}},\ }\href
  {https://doi.org/10.1103/PhysRevLett.105.135301} {\bibfield  {journal}
  {\bibinfo  {journal} {Physical Review Letters}\ }\textbf {\bibinfo {volume}
  {105}},\ \bibinfo {pages} {135301} (\bibinfo {year} {2010})}\BibitemShut
  {NoStop}%
\bibitem [{\citenamefont {Henkel}\ \emph {et~al.}(2010)\citenamefont {Henkel},
  \citenamefont {Nath},\ and\ \citenamefont
  {Pohl}}]{henkel_2010_threedimensional}%
  \BibitemOpen
  \bibfield  {author} {\bibinfo {author} {\bibfnamefont {N.}~\bibnamefont
  {Henkel}}, \bibinfo {author} {\bibfnamefont {R.}~\bibnamefont {Nath}},\ and\
  \bibinfo {author} {\bibfnamefont {T.}~\bibnamefont {Pohl}},\ }\href
  {https://doi.org/10.1103/PhysRevLett.104.195302} {\bibfield  {journal}
  {\bibinfo  {journal} {Physical Review Letters}\ }\textbf {\bibinfo {volume}
  {104}},\ \bibinfo {pages} {195302} (\bibinfo {year} {2010})}\BibitemShut
  {NoStop}%
\bibitem [{\citenamefont {Henkel}\ \emph {et~al.}(2012)\citenamefont {Henkel},
  \citenamefont {Cinti}, \citenamefont {Jain}, \citenamefont {Pupillo},\ and\
  \citenamefont {Pohl}}]{henkel_2012_supersolid}%
  \BibitemOpen
  \bibfield  {author} {\bibinfo {author} {\bibfnamefont {N.}~\bibnamefont
  {Henkel}}, \bibinfo {author} {\bibfnamefont {F.}~\bibnamefont {Cinti}},
  \bibinfo {author} {\bibfnamefont {P.}~\bibnamefont {Jain}}, \bibinfo {author}
  {\bibfnamefont {G.}~\bibnamefont {Pupillo}},\ and\ \bibinfo {author}
  {\bibfnamefont {T.}~\bibnamefont {Pohl}},\ }\href
  {https://doi.org/10.1103/PhysRevLett.108.265301} {\bibfield  {journal}
  {\bibinfo  {journal} {Physical Review Letters}\ }\textbf {\bibinfo {volume}
  {108}},\ \bibinfo {pages} {265301} (\bibinfo {year} {2012})}\BibitemShut
  {NoStop}%
\bibitem [{\citenamefont {Wang}\ \emph {et~al.}(2010)\citenamefont {Wang},
  \citenamefont {Gao}, \citenamefont {Jian},\ and\ \citenamefont
  {Zhai}}]{wang_2010_spinorbit}%
  \BibitemOpen
  \bibfield  {author} {\bibinfo {author} {\bibfnamefont {C.}~\bibnamefont
  {Wang}}, \bibinfo {author} {\bibfnamefont {C.}~\bibnamefont {Gao}}, \bibinfo
  {author} {\bibfnamefont {C.-M.}\ \bibnamefont {Jian}},\ and\ \bibinfo
  {author} {\bibfnamefont {H.}~\bibnamefont {Zhai}},\ }\href
  {https://doi.org/10.1103/PhysRevLett.105.160403} {\bibfield  {journal}
  {\bibinfo  {journal} {Physical Review Letters}\ }\textbf {\bibinfo {volume}
  {105}},\ \bibinfo {pages} {160403} (\bibinfo {year} {2010})}\BibitemShut
  {NoStop}%
\bibitem [{\citenamefont {Li}\ \emph {et~al.}(2013)\citenamefont {Li},
  \citenamefont {Martone}, \citenamefont {Pitaevskii},\ and\ \citenamefont
  {Stringari}}]{li_2013_superstripes}%
  \BibitemOpen
  \bibfield  {author} {\bibinfo {author} {\bibfnamefont {Y.}~\bibnamefont
  {Li}}, \bibinfo {author} {\bibfnamefont {G.~I.}\ \bibnamefont {Martone}},
  \bibinfo {author} {\bibfnamefont {L.~P.}\ \bibnamefont {Pitaevskii}},\ and\
  \bibinfo {author} {\bibfnamefont {S.}~\bibnamefont {Stringari}},\ }\href
  {https://doi.org/10.1103/PhysRevLett.110.235302} {\bibfield  {journal}
  {\bibinfo  {journal} {Physical Review Letters}\ }\textbf {\bibinfo {volume}
  {110}},\ \bibinfo {pages} {235302} (\bibinfo {year} {2013})}\BibitemShut
  {NoStop}%
\bibitem [{\citenamefont {Li}\ \emph {et~al.}(2016)\citenamefont {Li},
  \citenamefont {Huang}, \citenamefont {Shteynas}, \citenamefont {Burchesky},
  \citenamefont {Top}, \citenamefont {Su}, \citenamefont {Lee}, \citenamefont
  {Jamison},\ and\ \citenamefont {Ketterle}}]{li_2016_spinorbit}%
  \BibitemOpen
  \bibfield  {author} {\bibinfo {author} {\bibfnamefont {J.}~\bibnamefont
  {Li}}, \bibinfo {author} {\bibfnamefont {W.}~\bibnamefont {Huang}}, \bibinfo
  {author} {\bibfnamefont {B.}~\bibnamefont {Shteynas}}, \bibinfo {author}
  {\bibfnamefont {S.}~\bibnamefont {Burchesky}}, \bibinfo {author}
  {\bibfnamefont {F.~{\c C}.}\ \bibnamefont {Top}}, \bibinfo {author}
  {\bibfnamefont {E.}~\bibnamefont {Su}}, \bibinfo {author} {\bibfnamefont
  {J.}~\bibnamefont {Lee}}, \bibinfo {author} {\bibfnamefont {A.~O.}\
  \bibnamefont {Jamison}},\ and\ \bibinfo {author} {\bibfnamefont
  {W.}~\bibnamefont {Ketterle}},\ }\href
  {https://doi.org/10.1103/PhysRevLett.117.185301} {\bibfield  {journal}
  {\bibinfo  {journal} {Physical Review Letters}\ }\textbf {\bibinfo {volume}
  {117}},\ \bibinfo {pages} {185301} (\bibinfo {year} {2016})}\BibitemShut
  {NoStop}%
\bibitem [{\citenamefont {Li}\ \emph {et~al.}(2017)\citenamefont {Li},
  \citenamefont {Lee}, \citenamefont {Huang}, \citenamefont {Burchesky},
  \citenamefont {Shteynas}, \citenamefont {Top}, \citenamefont {Jamison},\ and\
  \citenamefont {Ketterle}}]{li_2017_stripe}%
  \BibitemOpen
  \bibfield  {author} {\bibinfo {author} {\bibfnamefont {J.-R.}\ \bibnamefont
  {Li}}, \bibinfo {author} {\bibfnamefont {J.}~\bibnamefont {Lee}}, \bibinfo
  {author} {\bibfnamefont {W.}~\bibnamefont {Huang}}, \bibinfo {author}
  {\bibfnamefont {S.}~\bibnamefont {Burchesky}}, \bibinfo {author}
  {\bibfnamefont {B.}~\bibnamefont {Shteynas}}, \bibinfo {author}
  {\bibfnamefont {F.~{\c C}.}\ \bibnamefont {Top}}, \bibinfo {author}
  {\bibfnamefont {A.~O.}\ \bibnamefont {Jamison}},\ and\ \bibinfo {author}
  {\bibfnamefont {W.}~\bibnamefont {Ketterle}},\ }\href
  {https://doi.org/10.1038/nature21431} {\bibfield  {journal} {\bibinfo
  {journal} {Nature}\ }\textbf {\bibinfo {volume} {543}},\ \bibinfo {pages}
  {91} (\bibinfo {year} {2017})}\BibitemShut {NoStop}%
\bibitem [{\citenamefont {Bersano}\ \emph {et~al.}(2019)\citenamefont
  {Bersano}, \citenamefont {Hou}, \citenamefont {Mossman}, \citenamefont
  {Gokhroo}, \citenamefont {Luo}, \citenamefont {Sun}, \citenamefont {Zhang},\
  and\ \citenamefont {Engels}}]{bersano_2019_experimental}%
  \BibitemOpen
  \bibfield  {author} {\bibinfo {author} {\bibfnamefont {T.~M.}\ \bibnamefont
  {Bersano}}, \bibinfo {author} {\bibfnamefont {J.}~\bibnamefont {Hou}},
  \bibinfo {author} {\bibfnamefont {S.}~\bibnamefont {Mossman}}, \bibinfo
  {author} {\bibfnamefont {V.}~\bibnamefont {Gokhroo}}, \bibinfo {author}
  {\bibfnamefont {X.-W.}\ \bibnamefont {Luo}}, \bibinfo {author} {\bibfnamefont
  {K.}~\bibnamefont {Sun}}, \bibinfo {author} {\bibfnamefont {C.}~\bibnamefont
  {Zhang}},\ and\ \bibinfo {author} {\bibfnamefont {P.}~\bibnamefont
  {Engels}},\ }\href {https://doi.org/10.1103/PhysRevA.99.051602} {\bibfield
  {journal} {\bibinfo  {journal} {Physical Review A}\ }\textbf {\bibinfo
  {volume} {99}},\ \bibinfo {pages} {051602} (\bibinfo {year}
  {2019})}\BibitemShut {NoStop}%
\bibitem [{\citenamefont {Putra}\ \emph {et~al.}(2020)\citenamefont {Putra},
  \citenamefont {{Salces-C{\'a}rcoba}}, \citenamefont {Yue}, \citenamefont
  {Sugawa},\ and\ \citenamefont {Spielman}}]{putra_2020_spatial}%
  \BibitemOpen
  \bibfield  {author} {\bibinfo {author} {\bibfnamefont {A.}~\bibnamefont
  {Putra}}, \bibinfo {author} {\bibfnamefont {F.}~\bibnamefont
  {{Salces-C{\'a}rcoba}}}, \bibinfo {author} {\bibfnamefont {Y.}~\bibnamefont
  {Yue}}, \bibinfo {author} {\bibfnamefont {S.}~\bibnamefont {Sugawa}},\ and\
  \bibinfo {author} {\bibfnamefont {I.~B.}\ \bibnamefont {Spielman}},\ }\href
  {https://doi.org/10.1103/PhysRevLett.124.053605} {\bibfield  {journal}
  {\bibinfo  {journal} {Physical Review Letters}\ }\textbf {\bibinfo {volume}
  {124}},\ \bibinfo {pages} {053605} (\bibinfo {year} {2020})}\BibitemShut
  {NoStop}%
\bibitem [{\citenamefont {Sachdeva}\ \emph {et~al.}(2020)\citenamefont
  {Sachdeva}, \citenamefont {Tengstrand},\ and\ \citenamefont
  {Reimann}}]{sachdeva_2020_selfbound}%
  \BibitemOpen
  \bibfield  {author} {\bibinfo {author} {\bibfnamefont {R.}~\bibnamefont
  {Sachdeva}}, \bibinfo {author} {\bibfnamefont {M.~N.}\ \bibnamefont
  {Tengstrand}},\ and\ \bibinfo {author} {\bibfnamefont {S.~M.}\ \bibnamefont
  {Reimann}},\ }\href {https://doi.org/10.1103/PhysRevA.102.043304} {\bibfield
  {journal} {\bibinfo  {journal} {Physical Review A}\ }\textbf {\bibinfo
  {volume} {102}},\ \bibinfo {pages} {043304} (\bibinfo {year}
  {2020})}\BibitemShut {NoStop}%
\bibitem [{\citenamefont {Geier}\ \emph {et~al.}(2021)\citenamefont {Geier},
  \citenamefont {Martone}, \citenamefont {Hauke},\ and\ \citenamefont
  {Stringari}}]{geier_2021_exciting}%
  \BibitemOpen
  \bibfield  {author} {\bibinfo {author} {\bibfnamefont {K.~T.}\ \bibnamefont
  {Geier}}, \bibinfo {author} {\bibfnamefont {G.~I.}\ \bibnamefont {Martone}},
  \bibinfo {author} {\bibfnamefont {P.}~\bibnamefont {Hauke}},\ and\ \bibinfo
  {author} {\bibfnamefont {S.}~\bibnamefont {Stringari}},\ }\href
  {https://doi.org/10.1103/PhysRevLett.127.115301} {\bibfield  {journal}
  {\bibinfo  {journal} {Physical Review Letters}\ }\textbf {\bibinfo {volume}
  {127}},\ \bibinfo {pages} {115301} (\bibinfo {year} {2021})}\BibitemShut
  {NoStop}%
\bibitem [{\citenamefont {L{\'e}onard}\ \emph {et~al.}(2017)\citenamefont
  {L{\'e}onard}, \citenamefont {Morales}, \citenamefont {Zupancic},
  \citenamefont {Esslinger},\ and\ \citenamefont
  {Donner}}]{leonard_2017_supersolid}%
  \BibitemOpen
  \bibfield  {author} {\bibinfo {author} {\bibfnamefont {J.}~\bibnamefont
  {L{\'e}onard}}, \bibinfo {author} {\bibfnamefont {A.}~\bibnamefont
  {Morales}}, \bibinfo {author} {\bibfnamefont {P.}~\bibnamefont {Zupancic}},
  \bibinfo {author} {\bibfnamefont {T.}~\bibnamefont {Esslinger}},\ and\
  \bibinfo {author} {\bibfnamefont {T.}~\bibnamefont {Donner}},\ }\href
  {https://doi.org/10.1038/nature21067} {\bibfield  {journal} {\bibinfo
  {journal} {Nature}\ }\textbf {\bibinfo {volume} {543}},\ \bibinfo {pages}
  {87} (\bibinfo {year} {2017})}\BibitemShut {NoStop}%
\bibitem [{\citenamefont {Zhang}\ \emph {et~al.}(2022)\citenamefont {Zhang},
  \citenamefont {Zhang}, \citenamefont {Yang},\ and\ \citenamefont
  {{Capogrosso-Sansone}}}]{zhang_2022_supersolid}%
  \BibitemOpen
  \bibfield  {author} {\bibinfo {author} {\bibfnamefont {J.}~\bibnamefont
  {Zhang}}, \bibinfo {author} {\bibfnamefont {C.}~\bibnamefont {Zhang}},
  \bibinfo {author} {\bibfnamefont {J.}~\bibnamefont {Yang}},\ and\ \bibinfo
  {author} {\bibfnamefont {B.}~\bibnamefont {{Capogrosso-Sansone}}},\ }\href
  {https://doi.org/10.1103/PhysRevA.105.063302} {\bibfield  {journal} {\bibinfo
   {journal} {Physical Review A}\ }\textbf {\bibinfo {volume} {105}},\ \bibinfo
  {pages} {063302} (\bibinfo {year} {2022})}\BibitemShut {NoStop}%
\bibitem [{\citenamefont {{Ferrier-Barbut}}\ \emph {et~al.}(2016)\citenamefont
  {{Ferrier-Barbut}}, \citenamefont {Kadau}, \citenamefont {Schmitt},
  \citenamefont {Wenzel},\ and\ \citenamefont
  {Pfau}}]{ferrier-barbut_2016_observation}%
  \BibitemOpen
  \bibfield  {author} {\bibinfo {author} {\bibfnamefont {I.}~\bibnamefont
  {{Ferrier-Barbut}}}, \bibinfo {author} {\bibfnamefont {H.}~\bibnamefont
  {Kadau}}, \bibinfo {author} {\bibfnamefont {M.}~\bibnamefont {Schmitt}},
  \bibinfo {author} {\bibfnamefont {M.}~\bibnamefont {Wenzel}},\ and\ \bibinfo
  {author} {\bibfnamefont {T.}~\bibnamefont {Pfau}},\ }\href
  {https://doi.org/10.1103/PhysRevLett.116.215301} {\bibfield  {journal}
  {\bibinfo  {journal} {Physical Review Letters}\ }\textbf {\bibinfo {volume}
  {116}},\ \bibinfo {pages} {215301} (\bibinfo {year} {2016})}\BibitemShut
  {NoStop}%
\bibitem [{\citenamefont {Kadau}\ \emph {et~al.}(2016)\citenamefont {Kadau},
  \citenamefont {Schmitt}, \citenamefont {Wenzel}, \citenamefont {Wink},
  \citenamefont {Maier}, \citenamefont {{Ferrier-Barbut}},\ and\ \citenamefont
  {Pfau}}]{kadau_2016_observing}%
  \BibitemOpen
  \bibfield  {author} {\bibinfo {author} {\bibfnamefont {H.}~\bibnamefont
  {Kadau}}, \bibinfo {author} {\bibfnamefont {M.}~\bibnamefont {Schmitt}},
  \bibinfo {author} {\bibfnamefont {M.}~\bibnamefont {Wenzel}}, \bibinfo
  {author} {\bibfnamefont {C.}~\bibnamefont {Wink}}, \bibinfo {author}
  {\bibfnamefont {T.}~\bibnamefont {Maier}}, \bibinfo {author} {\bibfnamefont
  {I.}~\bibnamefont {{Ferrier-Barbut}}},\ and\ \bibinfo {author} {\bibfnamefont
  {T.}~\bibnamefont {Pfau}},\ }\href {https://doi.org/10.1038/nature16485}
  {\bibfield  {journal} {\bibinfo  {journal} {Nature}\ }\textbf {\bibinfo
  {volume} {530}},\ \bibinfo {pages} {194} (\bibinfo {year}
  {2016})}\BibitemShut {NoStop}%
\bibitem [{\citenamefont {Schmitt}\ \emph {et~al.}(2016)\citenamefont
  {Schmitt}, \citenamefont {Wenzel}, \citenamefont {B{\"o}ttcher},
  \citenamefont {{Ferrier-Barbut}},\ and\ \citenamefont
  {Pfau}}]{schmitt_2016_selfbound}%
  \BibitemOpen
  \bibfield  {author} {\bibinfo {author} {\bibfnamefont {M.}~\bibnamefont
  {Schmitt}}, \bibinfo {author} {\bibfnamefont {M.}~\bibnamefont {Wenzel}},
  \bibinfo {author} {\bibfnamefont {F.}~\bibnamefont {B{\"o}ttcher}}, \bibinfo
  {author} {\bibfnamefont {I.}~\bibnamefont {{Ferrier-Barbut}}},\ and\ \bibinfo
  {author} {\bibfnamefont {T.}~\bibnamefont {Pfau}},\ }\href
  {https://doi.org/10.1038/nature20126} {\bibfield  {journal} {\bibinfo
  {journal} {Nature}\ }\textbf {\bibinfo {volume} {539}},\ \bibinfo {pages}
  {259} (\bibinfo {year} {2016})}\BibitemShut {NoStop}%
\bibitem [{\citenamefont {Ilzh{\"o}fer}\ \emph {et~al.}(2021)\citenamefont
  {Ilzh{\"o}fer}, \citenamefont {Sohmen}, \citenamefont {Durastante},
  \citenamefont {Politi}, \citenamefont {Trautmann}, \citenamefont {Natale},
  \citenamefont {Morpurgo}, \citenamefont {Giamarchi}, \citenamefont {Chomaz},
  \citenamefont {Mark},\ and\ \citenamefont {Ferlaino}}]{ilzhofer_2021_phase}%
  \BibitemOpen
  \bibfield  {author} {\bibinfo {author} {\bibfnamefont {P.}~\bibnamefont
  {Ilzh{\"o}fer}}, \bibinfo {author} {\bibfnamefont {M.}~\bibnamefont
  {Sohmen}}, \bibinfo {author} {\bibfnamefont {G.}~\bibnamefont {Durastante}},
  \bibinfo {author} {\bibfnamefont {C.}~\bibnamefont {Politi}}, \bibinfo
  {author} {\bibfnamefont {A.}~\bibnamefont {Trautmann}}, \bibinfo {author}
  {\bibfnamefont {G.}~\bibnamefont {Natale}}, \bibinfo {author} {\bibfnamefont
  {G.}~\bibnamefont {Morpurgo}}, \bibinfo {author} {\bibfnamefont
  {T.}~\bibnamefont {Giamarchi}}, \bibinfo {author} {\bibfnamefont
  {L.}~\bibnamefont {Chomaz}}, \bibinfo {author} {\bibfnamefont {M.~J.}\
  \bibnamefont {Mark}},\ and\ \bibinfo {author} {\bibfnamefont
  {F.}~\bibnamefont {Ferlaino}},\ }\href
  {https://doi.org/10.1038/s41567-020-01100-3} {\bibfield  {journal} {\bibinfo
  {journal} {Nature Physics}\ }\textbf {\bibinfo {volume} {17}},\ \bibinfo
  {pages} {356} (\bibinfo {year} {2021})}\BibitemShut {NoStop}%
\bibitem [{\citenamefont {Tang}\ \emph
  {et~al.}(2015{\natexlab{a}})\citenamefont {Tang}, \citenamefont {Sykes},
  \citenamefont {Burdick}, \citenamefont {Bohn},\ and\ \citenamefont
  {Lev}}]{tang_2015_wave}%
  \BibitemOpen
  \bibfield  {author} {\bibinfo {author} {\bibfnamefont {Y.}~\bibnamefont
  {Tang}}, \bibinfo {author} {\bibfnamefont {A.}~\bibnamefont {Sykes}},
  \bibinfo {author} {\bibfnamefont {N.~Q.}\ \bibnamefont {Burdick}}, \bibinfo
  {author} {\bibfnamefont {J.~L.}\ \bibnamefont {Bohn}},\ and\ \bibinfo
  {author} {\bibfnamefont {B.~L.}\ \bibnamefont {Lev}},\ }\href
  {https://doi.org/10.1103/PhysRevA.92.022703} {\bibfield  {journal} {\bibinfo
  {journal} {Physical Review A}\ }\textbf {\bibinfo {volume} {92}},\ \bibinfo
  {pages} {022703} (\bibinfo {year} {2015}{\natexlab{a}})}\BibitemShut
  {NoStop}%
\bibitem [{\citenamefont {Tang}\ \emph
  {et~al.}(2015{\natexlab{b}})\citenamefont {Tang}, \citenamefont {Burdick},
  \citenamefont {Baumann},\ and\ \citenamefont {Lev}}]{tang_2015_bose}%
  \BibitemOpen
  \bibfield  {author} {\bibinfo {author} {\bibfnamefont {Y.}~\bibnamefont
  {Tang}}, \bibinfo {author} {\bibfnamefont {N.~Q.}\ \bibnamefont {Burdick}},
  \bibinfo {author} {\bibfnamefont {K.}~\bibnamefont {Baumann}},\ and\ \bibinfo
  {author} {\bibfnamefont {B.~L.}\ \bibnamefont {Lev}},\ }\href
  {https://doi.org/10.1088/1367-2630/17/4/045006} {\bibfield  {journal}
  {\bibinfo  {journal} {New Journal of Physics}\ }\textbf {\bibinfo {volume}
  {17}},\ \bibinfo {pages} {045006} (\bibinfo {year}
  {2015}{\natexlab{b}})}\BibitemShut {NoStop}%
\bibitem [{\citenamefont {Chomaz}\ \emph {et~al.}(2016)\citenamefont {Chomaz},
  \citenamefont {Baier}, \citenamefont {Petter}, \citenamefont {Mark},
  \citenamefont {W\"achtler}, \citenamefont {Santos},\ and\ \citenamefont
  {Ferlaino}}]{chomaz_2016_quantumfluctuationdriven}%
  \BibitemOpen
  \bibfield  {author} {\bibinfo {author} {\bibfnamefont {L.}~\bibnamefont
  {Chomaz}}, \bibinfo {author} {\bibfnamefont {S.}~\bibnamefont {Baier}},
  \bibinfo {author} {\bibfnamefont {D.}~\bibnamefont {Petter}}, \bibinfo
  {author} {\bibfnamefont {M.~J.}\ \bibnamefont {Mark}}, \bibinfo {author}
  {\bibfnamefont {F.}~\bibnamefont {W\"achtler}}, \bibinfo {author}
  {\bibfnamefont {L.}~\bibnamefont {Santos}},\ and\ \bibinfo {author}
  {\bibfnamefont {F.}~\bibnamefont {Ferlaino}},\ }\href
  {https://doi.org/10.1103/PhysRevX.6.041039} {\bibfield  {journal} {\bibinfo
  {journal} {Phys. Rev. X}\ }\textbf {\bibinfo {volume} {6}},\ \bibinfo {pages}
  {041039} (\bibinfo {year} {2016})}\BibitemShut {NoStop}%
\bibitem [{\citenamefont {Chomaz}\ \emph {et~al.}(2019)\citenamefont {Chomaz},
  \citenamefont {Petter}, \citenamefont {Ilzh{\"o}fer}, \citenamefont {Natale},
  \citenamefont {Trautmann}, \citenamefont {Politi}, \citenamefont
  {Durastante}, \citenamefont {{van Bijnen}}, \citenamefont {Patscheider},
  \citenamefont {Sohmen}, \citenamefont {Mark},\ and\ \citenamefont
  {Ferlaino}}]{chomaz_2019_longlived}%
  \BibitemOpen
  \bibfield  {author} {\bibinfo {author} {\bibfnamefont {L.}~\bibnamefont
  {Chomaz}}, \bibinfo {author} {\bibfnamefont {D.}~\bibnamefont {Petter}},
  \bibinfo {author} {\bibfnamefont {P.}~\bibnamefont {Ilzh{\"o}fer}}, \bibinfo
  {author} {\bibfnamefont {G.}~\bibnamefont {Natale}}, \bibinfo {author}
  {\bibfnamefont {A.}~\bibnamefont {Trautmann}}, \bibinfo {author}
  {\bibfnamefont {C.}~\bibnamefont {Politi}}, \bibinfo {author} {\bibfnamefont
  {G.}~\bibnamefont {Durastante}}, \bibinfo {author} {\bibfnamefont {R.~M.~W.}\
  \bibnamefont {{van Bijnen}}}, \bibinfo {author} {\bibfnamefont
  {A.}~\bibnamefont {Patscheider}}, \bibinfo {author} {\bibfnamefont
  {M.}~\bibnamefont {Sohmen}}, \bibinfo {author} {\bibfnamefont {M.~J.}\
  \bibnamefont {Mark}},\ and\ \bibinfo {author} {\bibfnamefont
  {F.}~\bibnamefont {Ferlaino}},\ }\href
  {https://doi.org/10.1103/PhysRevX.9.021012} {\bibfield  {journal} {\bibinfo
  {journal} {Physical Review X}\ }\textbf {\bibinfo {volume} {9}},\ \bibinfo
  {pages} {021012} (\bibinfo {year} {2019})}\BibitemShut {NoStop}%
\bibitem [{\citenamefont {Natale}\ \emph {et~al.}(2019)\citenamefont {Natale},
  \citenamefont {{van Bijnen}}, \citenamefont {Patscheider}, \citenamefont
  {Petter}, \citenamefont {Mark}, \citenamefont {Chomaz},\ and\ \citenamefont
  {Ferlaino}}]{natale_2019_excitation}%
  \BibitemOpen
  \bibfield  {author} {\bibinfo {author} {\bibfnamefont {G.}~\bibnamefont
  {Natale}}, \bibinfo {author} {\bibfnamefont {R.~M.~W.}\ \bibnamefont {{van
  Bijnen}}}, \bibinfo {author} {\bibfnamefont {A.}~\bibnamefont {Patscheider}},
  \bibinfo {author} {\bibfnamefont {D.}~\bibnamefont {Petter}}, \bibinfo
  {author} {\bibfnamefont {M.~J.}\ \bibnamefont {Mark}}, \bibinfo {author}
  {\bibfnamefont {L.}~\bibnamefont {Chomaz}},\ and\ \bibinfo {author}
  {\bibfnamefont {F.}~\bibnamefont {Ferlaino}},\ }\href
  {https://doi.org/10.1103/PhysRevLett.123.050402} {\bibfield  {journal}
  {\bibinfo  {journal} {Physical Review Letters}\ }\textbf {\bibinfo {volume}
  {123}},\ \bibinfo {pages} {050402} (\bibinfo {year} {2019})}\BibitemShut
  {NoStop}%
\bibitem [{\citenamefont {Lima}\ and\ \citenamefont
  {Pelster}(2011)}]{lima_2011_quantum}%
  \BibitemOpen
  \bibfield  {author} {\bibinfo {author} {\bibfnamefont {A.~R.~P.}\
  \bibnamefont {Lima}}\ and\ \bibinfo {author} {\bibfnamefont {A.}~\bibnamefont
  {Pelster}},\ }\href {https://doi.org/10.1103/PhysRevA.84.041604} {\bibfield
  {journal} {\bibinfo  {journal} {Phys. Rev. A}\ }\textbf {\bibinfo {volume}
  {84}},\ \bibinfo {pages} {041604} (\bibinfo {year} {2011})}\BibitemShut
  {NoStop}%
\bibitem [{\citenamefont {Lima}\ and\ \citenamefont
  {Pelster}(2012)}]{lima_2012_meanfield}%
  \BibitemOpen
  \bibfield  {author} {\bibinfo {author} {\bibfnamefont {A.~R.~P.}\
  \bibnamefont {Lima}}\ and\ \bibinfo {author} {\bibfnamefont {A.}~\bibnamefont
  {Pelster}},\ }\href {https://doi.org/10.1103/PhysRevA.86.063609} {\bibfield
  {journal} {\bibinfo  {journal} {Phys. Rev. A}\ }\textbf {\bibinfo {volume}
  {86}},\ \bibinfo {pages} {063609} (\bibinfo {year} {2012})}\BibitemShut
  {NoStop}%
\bibitem [{\citenamefont {Petrov}(2015)}]{petrov_2015_quantum}%
  \BibitemOpen
  \bibfield  {author} {\bibinfo {author} {\bibfnamefont {D.~S.}\ \bibnamefont
  {Petrov}},\ }\href {https://doi.org/10.1103/PhysRevLett.115.155302}
  {\bibfield  {journal} {\bibinfo  {journal} {Physical Review Letters}\
  }\textbf {\bibinfo {volume} {115}},\ \bibinfo {pages} {155302} (\bibinfo
  {year} {2015})}\BibitemShut {NoStop}%
\bibitem [{\citenamefont {Mulkerin}\ \emph {et~al.}(2013)\citenamefont
  {Mulkerin}, \citenamefont {Van~Bijnen}, \citenamefont {O'Dell}, \citenamefont
  {Martin},\ and\ \citenamefont {Parker}}]{mulkerin_2013_anisotropic}%
  \BibitemOpen
  \bibfield  {author} {\bibinfo {author} {\bibfnamefont {B.~C.}\ \bibnamefont
  {Mulkerin}}, \bibinfo {author} {\bibfnamefont {R.~M.~W.}\ \bibnamefont
  {Van~Bijnen}}, \bibinfo {author} {\bibfnamefont {D.~H.~J.}\ \bibnamefont
  {O'Dell}}, \bibinfo {author} {\bibfnamefont {A.~M.}\ \bibnamefont {Martin}},\
  and\ \bibinfo {author} {\bibfnamefont {N.~G.}\ \bibnamefont {Parker}},\
  }\href {https://doi.org/10.1103/PhysRevLett.111.170402} {\bibfield  {journal}
  {\bibinfo  {journal} {Physical Review Letters}\ }\textbf {\bibinfo {volume}
  {111}},\ \bibinfo {pages} {170402} (\bibinfo {year} {2013})}\BibitemShut
  {NoStop}%
\bibitem [{\citenamefont {Martin}\ \emph {et~al.}(2017)\citenamefont {Martin},
  \citenamefont {Marchant}, \citenamefont {O'Dell},\ and\ \citenamefont
  {Parker}}]{martin_2017_vortices}%
  \BibitemOpen
  \bibfield  {author} {\bibinfo {author} {\bibfnamefont {A.~M.}\ \bibnamefont
  {Martin}}, \bibinfo {author} {\bibfnamefont {N.~G.}\ \bibnamefont
  {Marchant}}, \bibinfo {author} {\bibfnamefont {D.~H.~J.}\ \bibnamefont
  {O'Dell}},\ and\ \bibinfo {author} {\bibfnamefont {N.~G.}\ \bibnamefont
  {Parker}},\ }\href {https://doi.org/10.1088/1361-648X/aa53a6} {\bibfield
  {journal} {\bibinfo  {journal} {Journal of Physics: Condensed Matter}\
  }\textbf {\bibinfo {volume} {29}},\ \bibinfo {pages} {103004} (\bibinfo
  {year} {2017})}\BibitemShut {NoStop}%
\bibitem [{\citenamefont {Santos}\ \emph {et~al.}(2003)\citenamefont {Santos},
  \citenamefont {Shlyapnikov},\ and\ \citenamefont
  {Lewenstein}}]{santos_2003_rotonmaxon}%
  \BibitemOpen
  \bibfield  {author} {\bibinfo {author} {\bibfnamefont {L.}~\bibnamefont
  {Santos}}, \bibinfo {author} {\bibfnamefont {G.~V.}\ \bibnamefont
  {Shlyapnikov}},\ and\ \bibinfo {author} {\bibfnamefont {M.}~\bibnamefont
  {Lewenstein}},\ }\href {https://doi.org/10.1103/PhysRevLett.90.250403}
  {\bibfield  {journal} {\bibinfo  {journal} {Physical Review Letters}\
  }\textbf {\bibinfo {volume} {90}},\ \bibinfo {pages} {250403} (\bibinfo
  {year} {2003})}\BibitemShut {NoStop}%
\bibitem [{\citenamefont {Bisset}\ \emph {et~al.}(2013)\citenamefont {Bisset},
  \citenamefont {Baillie},\ and\ \citenamefont {Blakie}}]{bisset_2013_roton}%
  \BibitemOpen
  \bibfield  {author} {\bibinfo {author} {\bibfnamefont {R.~N.}\ \bibnamefont
  {Bisset}}, \bibinfo {author} {\bibfnamefont {D.}~\bibnamefont {Baillie}},\
  and\ \bibinfo {author} {\bibfnamefont {P.~B.}\ \bibnamefont {Blakie}},\
  }\href {https://doi.org/10.1103/PhysRevA.88.043606} {\bibfield  {journal}
  {\bibinfo  {journal} {Physical Review A}\ }\textbf {\bibinfo {volume} {88}},\
  \bibinfo {pages} {043606} (\bibinfo {year} {2013})}\BibitemShut {NoStop}%
\bibitem [{\citenamefont {Baillie}\ \emph {et~al.}(2017)\citenamefont
  {Baillie}, \citenamefont {Wilson},\ and\ \citenamefont
  {Blakie}}]{baillie_2017_collective}%
  \BibitemOpen
  \bibfield  {author} {\bibinfo {author} {\bibfnamefont {D.}~\bibnamefont
  {Baillie}}, \bibinfo {author} {\bibfnamefont {R.~M.}\ \bibnamefont
  {Wilson}},\ and\ \bibinfo {author} {\bibfnamefont {P.~B.}\ \bibnamefont
  {Blakie}},\ }\href {https://doi.org/10.1103/PhysRevLett.119.255302}
  {\bibfield  {journal} {\bibinfo  {journal} {Physical Review Letters}\
  }\textbf {\bibinfo {volume} {119}},\ \bibinfo {pages} {255302} (\bibinfo
  {year} {2017})}\BibitemShut {NoStop}%
\bibitem [{\citenamefont {Chomaz}\ \emph {et~al.}(2018)\citenamefont {Chomaz},
  \citenamefont {{van Bijnen}}, \citenamefont {Petter}, \citenamefont
  {Faraoni}, \citenamefont {Baier}, \citenamefont {Becher}, \citenamefont
  {Mark}, \citenamefont {W{\"a}chtler}, \citenamefont {Santos},\ and\
  \citenamefont {Ferlaino}}]{chomaz_2018_observation}%
  \BibitemOpen
  \bibfield  {author} {\bibinfo {author} {\bibfnamefont {L.}~\bibnamefont
  {Chomaz}}, \bibinfo {author} {\bibfnamefont {R.~M.~W.}\ \bibnamefont {{van
  Bijnen}}}, \bibinfo {author} {\bibfnamefont {D.}~\bibnamefont {Petter}},
  \bibinfo {author} {\bibfnamefont {G.}~\bibnamefont {Faraoni}}, \bibinfo
  {author} {\bibfnamefont {S.}~\bibnamefont {Baier}}, \bibinfo {author}
  {\bibfnamefont {J.~H.}\ \bibnamefont {Becher}}, \bibinfo {author}
  {\bibfnamefont {M.~J.}\ \bibnamefont {Mark}}, \bibinfo {author}
  {\bibfnamefont {F.}~\bibnamefont {W{\"a}chtler}}, \bibinfo {author}
  {\bibfnamefont {L.}~\bibnamefont {Santos}},\ and\ \bibinfo {author}
  {\bibfnamefont {F.}~\bibnamefont {Ferlaino}},\ }\href
  {https://doi.org/10.1038/s41567-018-0054-7} {\bibfield  {journal} {\bibinfo
  {journal} {Nature Physics}\ }\textbf {\bibinfo {volume} {14}},\ \bibinfo
  {pages} {442} (\bibinfo {year} {2018})}\BibitemShut {NoStop}%
\bibitem [{\citenamefont {Schmidt}\ \emph {et~al.}(2021)\citenamefont
  {Schmidt}, \citenamefont {Hertkorn}, \citenamefont {Guo}, \citenamefont
  {B{\"o}ttcher}, \citenamefont {Schmidt}, \citenamefont {Ng}, \citenamefont
  {Graham}, \citenamefont {Langen}, \citenamefont {Zwierlein},\ and\
  \citenamefont {Pfau}}]{schmidt_2021_roton}%
  \BibitemOpen
  \bibfield  {author} {\bibinfo {author} {\bibfnamefont {J.-N.}\ \bibnamefont
  {Schmidt}}, \bibinfo {author} {\bibfnamefont {J.}~\bibnamefont {Hertkorn}},
  \bibinfo {author} {\bibfnamefont {M.}~\bibnamefont {Guo}}, \bibinfo {author}
  {\bibfnamefont {F.}~\bibnamefont {B{\"o}ttcher}}, \bibinfo {author}
  {\bibfnamefont {M.}~\bibnamefont {Schmidt}}, \bibinfo {author} {\bibfnamefont
  {K.~S.~H.}\ \bibnamefont {Ng}}, \bibinfo {author} {\bibfnamefont {S.~D.}\
  \bibnamefont {Graham}}, \bibinfo {author} {\bibfnamefont {T.}~\bibnamefont
  {Langen}}, \bibinfo {author} {\bibfnamefont {M.}~\bibnamefont {Zwierlein}},\
  and\ \bibinfo {author} {\bibfnamefont {T.}~\bibnamefont {Pfau}},\ }\href
  {https://doi.org/10.1103/PhysRevLett.126.193002} {\bibfield  {journal}
  {\bibinfo  {journal} {Physical Review Letters}\ }\textbf {\bibinfo {volume}
  {126}},\ \bibinfo {pages} {193002} (\bibinfo {year} {2021})}\BibitemShut
  {NoStop}%
\bibitem [{\citenamefont {W{\"a}chtler}\ and\ \citenamefont
  {Santos}(2016{\natexlab{a}})}]{wachtler_2016_quantum}%
  \BibitemOpen
  \bibfield  {author} {\bibinfo {author} {\bibfnamefont {F.}~\bibnamefont
  {W{\"a}chtler}}\ and\ \bibinfo {author} {\bibfnamefont {L.}~\bibnamefont
  {Santos}},\ }\href {https://doi.org/10.1103/PhysRevA.93.061603} {\bibfield
  {journal} {\bibinfo  {journal} {Physical Review A}\ }\textbf {\bibinfo
  {volume} {93}},\ \bibinfo {pages} {061603} (\bibinfo {year}
  {2016}{\natexlab{a}})}\BibitemShut {NoStop}%
\bibitem [{\citenamefont {W{\"a}chtler}\ and\ \citenamefont
  {Santos}(2016{\natexlab{b}})}]{wachtler_2016_groundstate}%
  \BibitemOpen
  \bibfield  {author} {\bibinfo {author} {\bibfnamefont {F.}~\bibnamefont
  {W{\"a}chtler}}\ and\ \bibinfo {author} {\bibfnamefont {L.}~\bibnamefont
  {Santos}},\ }\href {https://doi.org/10.1103/PhysRevA.94.043618} {\bibfield
  {journal} {\bibinfo  {journal} {Physical Review A}\ }\textbf {\bibinfo
  {volume} {94}},\ \bibinfo {pages} {043618} (\bibinfo {year}
  {2016}{\natexlab{b}})}\BibitemShut {NoStop}%
\bibitem [{\citenamefont {Baillie}\ and\ \citenamefont
  {Blakie}(2018)}]{baillie_2018_droplet}%
  \BibitemOpen
  \bibfield  {author} {\bibinfo {author} {\bibfnamefont {D.}~\bibnamefont
  {Baillie}}\ and\ \bibinfo {author} {\bibfnamefont {P.~B.}\ \bibnamefont
  {Blakie}},\ }\href {https://doi.org/10.1103/PhysRevLett.121.195301}
  {\bibfield  {journal} {\bibinfo  {journal} {Physical Review Letters}\
  }\textbf {\bibinfo {volume} {121}},\ \bibinfo {pages} {195301} (\bibinfo
  {year} {2018})}\BibitemShut {NoStop}%
\bibitem [{\citenamefont {Mishra}\ \emph {et~al.}(2020)\citenamefont {Mishra},
  \citenamefont {Santos},\ and\ \citenamefont {Nath}}]{mishra_2020_selfbound}%
  \BibitemOpen
  \bibfield  {author} {\bibinfo {author} {\bibfnamefont {C.}~\bibnamefont
  {Mishra}}, \bibinfo {author} {\bibfnamefont {L.}~\bibnamefont {Santos}},\
  and\ \bibinfo {author} {\bibfnamefont {R.}~\bibnamefont {Nath}},\ }\href
  {https://doi.org/10.1103/PhysRevLett.124.073402} {\bibfield  {journal}
  {\bibinfo  {journal} {Physical Review Letters}\ }\textbf {\bibinfo {volume}
  {124}},\ \bibinfo {pages} {073402} (\bibinfo {year} {2020})},\ \Eprint
  {https://arxiv.org/abs/1907.08190} {arxiv:1907.08190} \BibitemShut {NoStop}%
\bibitem [{\citenamefont {Ghosh}\ \emph {et~al.}(2022)\citenamefont {Ghosh},
  \citenamefont {Mishra}, \citenamefont {Santos},\ and\ \citenamefont
  {Nath}}]{ghosh_2022_droplet}%
  \BibitemOpen
  \bibfield  {author} {\bibinfo {author} {\bibfnamefont {R.}~\bibnamefont
  {Ghosh}}, \bibinfo {author} {\bibfnamefont {C.}~\bibnamefont {Mishra}},
  \bibinfo {author} {\bibfnamefont {L.}~\bibnamefont {Santos}},\ and\ \bibinfo
  {author} {\bibfnamefont {R.}~\bibnamefont {Nath}},\ }\href
  {https://doi.org/10.1103/PhysRevA.106.063318} {\bibfield  {journal} {\bibinfo
   {journal} {Physical Review A}\ }\textbf {\bibinfo {volume} {106}},\ \bibinfo
  {pages} {063318} (\bibinfo {year} {2022})}\BibitemShut {NoStop}%
\bibitem [{\citenamefont {B{\"o}ttcher}\ \emph {et~al.}(2019)\citenamefont
  {B{\"o}ttcher}, \citenamefont {Schmidt}, \citenamefont {Wenzel},
  \citenamefont {Hertkorn}, \citenamefont {Guo}, \citenamefont {Langen},\ and\
  \citenamefont {Pfau}}]{bottcher_2019_transient}%
  \BibitemOpen
  \bibfield  {author} {\bibinfo {author} {\bibfnamefont {F.}~\bibnamefont
  {B{\"o}ttcher}}, \bibinfo {author} {\bibfnamefont {J.-N.}\ \bibnamefont
  {Schmidt}}, \bibinfo {author} {\bibfnamefont {M.}~\bibnamefont {Wenzel}},
  \bibinfo {author} {\bibfnamefont {J.}~\bibnamefont {Hertkorn}}, \bibinfo
  {author} {\bibfnamefont {M.}~\bibnamefont {Guo}}, \bibinfo {author}
  {\bibfnamefont {T.}~\bibnamefont {Langen}},\ and\ \bibinfo {author}
  {\bibfnamefont {T.}~\bibnamefont {Pfau}},\ }\href
  {https://doi.org/10.1103/PhysRevX.9.011051} {\bibfield  {journal} {\bibinfo
  {journal} {Physical Review X}\ }\textbf {\bibinfo {volume} {9}},\ \bibinfo
  {pages} {011051} (\bibinfo {year} {2019})}\BibitemShut {NoStop}%
\bibitem [{\citenamefont {Tanzi}\ \emph {et~al.}(2019)\citenamefont {Tanzi},
  \citenamefont {Lucioni}, \citenamefont {Fam{\`a}}, \citenamefont {Catani},
  \citenamefont {Fioretti}, \citenamefont {Gabbanini}, \citenamefont {Bisset},
  \citenamefont {Santos},\ and\ \citenamefont
  {Modugno}}]{tanzi_2019_observation}%
  \BibitemOpen
  \bibfield  {author} {\bibinfo {author} {\bibfnamefont {L.}~\bibnamefont
  {Tanzi}}, \bibinfo {author} {\bibfnamefont {E.}~\bibnamefont {Lucioni}},
  \bibinfo {author} {\bibfnamefont {F.}~\bibnamefont {Fam{\`a}}}, \bibinfo
  {author} {\bibfnamefont {J.}~\bibnamefont {Catani}}, \bibinfo {author}
  {\bibfnamefont {A.}~\bibnamefont {Fioretti}}, \bibinfo {author}
  {\bibfnamefont {C.}~\bibnamefont {Gabbanini}}, \bibinfo {author}
  {\bibfnamefont {R.~N.}\ \bibnamefont {Bisset}}, \bibinfo {author}
  {\bibfnamefont {L.}~\bibnamefont {Santos}},\ and\ \bibinfo {author}
  {\bibfnamefont {G.}~\bibnamefont {Modugno}},\ }\href
  {https://doi.org/10.1103/PhysRevLett.122.130405} {\bibfield  {journal}
  {\bibinfo  {journal} {Physical Review Letters}\ }\textbf {\bibinfo {volume}
  {122}},\ \bibinfo {pages} {130405} (\bibinfo {year} {2019})}\BibitemShut
  {NoStop}%
\bibitem [{\citenamefont {Roccuzzo}\ and\ \citenamefont
  {Ancilotto}(2019)}]{roccuzzo_2019_supersolid}%
  \BibitemOpen
  \bibfield  {author} {\bibinfo {author} {\bibfnamefont {S.~M.}\ \bibnamefont
  {Roccuzzo}}\ and\ \bibinfo {author} {\bibfnamefont {F.}~\bibnamefont
  {Ancilotto}},\ }\href {https://doi.org/10.1103/PhysRevA.99.041601} {\bibfield
   {journal} {\bibinfo  {journal} {Physical Review A}\ }\textbf {\bibinfo
  {volume} {99}},\ \bibinfo {pages} {041601} (\bibinfo {year}
  {2019})}\BibitemShut {NoStop}%
\bibitem [{\citenamefont {Blakie}\ \emph {et~al.}(2020)\citenamefont {Blakie},
  \citenamefont {Baillie}, \citenamefont {Chomaz},\ and\ \citenamefont
  {Ferlaino}}]{blakie_2020_supersolidity}%
  \BibitemOpen
  \bibfield  {author} {\bibinfo {author} {\bibfnamefont {P.~B.}\ \bibnamefont
  {Blakie}}, \bibinfo {author} {\bibfnamefont {D.}~\bibnamefont {Baillie}},
  \bibinfo {author} {\bibfnamefont {L.}~\bibnamefont {Chomaz}},\ and\ \bibinfo
  {author} {\bibfnamefont {F.}~\bibnamefont {Ferlaino}},\ }\href
  {https://doi.org/10.1103/PhysRevResearch.2.043318} {\bibfield  {journal}
  {\bibinfo  {journal} {Physical Review Research}\ }\textbf {\bibinfo {volume}
  {2}},\ \bibinfo {pages} {043318} (\bibinfo {year} {2020})}\BibitemShut
  {NoStop}%
\bibitem [{\citenamefont {Smith}\ \emph {et~al.}(2023)\citenamefont {Smith},
  \citenamefont {Baillie},\ and\ \citenamefont
  {Blakie}}]{smith_2023_supersolidity}%
  \BibitemOpen
  \bibfield  {author} {\bibinfo {author} {\bibfnamefont {J.~C.}\ \bibnamefont
  {Smith}}, \bibinfo {author} {\bibfnamefont {D.}~\bibnamefont {Baillie}},\
  and\ \bibinfo {author} {\bibfnamefont {P.~B.}\ \bibnamefont {Blakie}},\
  }\href {https://doi.org/10.1103/PhysRevA.107.033301} {\bibfield  {journal}
  {\bibinfo  {journal} {Physical Review A}\ }\textbf {\bibinfo {volume}
  {107}},\ \bibinfo {pages} {033301} (\bibinfo {year} {2023})}\BibitemShut
  {NoStop}%
\bibitem [{\citenamefont {{S{\'a}nchez-Baena}}\ \emph
  {et~al.}(2023)\citenamefont {{S{\'a}nchez-Baena}}, \citenamefont {Politi},
  \citenamefont {Maucher}, \citenamefont {Ferlaino},\ and\ \citenamefont
  {Pohl}}]{sanchez-baena_2023_heating}%
  \BibitemOpen
  \bibfield  {author} {\bibinfo {author} {\bibfnamefont {J.}~\bibnamefont
  {{S{\'a}nchez-Baena}}}, \bibinfo {author} {\bibfnamefont {C.}~\bibnamefont
  {Politi}}, \bibinfo {author} {\bibfnamefont {F.}~\bibnamefont {Maucher}},
  \bibinfo {author} {\bibfnamefont {F.}~\bibnamefont {Ferlaino}},\ and\
  \bibinfo {author} {\bibfnamefont {T.}~\bibnamefont {Pohl}},\ }\href
  {https://doi.org/10.1038/s41467-023-37207-3} {\bibfield  {journal} {\bibinfo
  {journal} {Nature Communications}\ }\textbf {\bibinfo {volume} {14}},\
  \bibinfo {pages} {1868} (\bibinfo {year} {2023})}\BibitemShut {NoStop}%
\bibitem [{\citenamefont {Norcia}\ \emph {et~al.}(2021)\citenamefont {Norcia},
  \citenamefont {Politi}, \citenamefont {Klaus}, \citenamefont {Poli},
  \citenamefont {Sohmen}, \citenamefont {Mark}, \citenamefont {Bisset},
  \citenamefont {Santos},\ and\ \citenamefont
  {Ferlaino}}]{norcia_2021_twodimensional}%
  \BibitemOpen
  \bibfield  {author} {\bibinfo {author} {\bibfnamefont {M.~A.}\ \bibnamefont
  {Norcia}}, \bibinfo {author} {\bibfnamefont {C.}~\bibnamefont {Politi}},
  \bibinfo {author} {\bibfnamefont {L.}~\bibnamefont {Klaus}}, \bibinfo
  {author} {\bibfnamefont {E.}~\bibnamefont {Poli}}, \bibinfo {author}
  {\bibfnamefont {M.}~\bibnamefont {Sohmen}}, \bibinfo {author} {\bibfnamefont
  {M.~J.}\ \bibnamefont {Mark}}, \bibinfo {author} {\bibfnamefont {R.~N.}\
  \bibnamefont {Bisset}}, \bibinfo {author} {\bibfnamefont {L.}~\bibnamefont
  {Santos}},\ and\ \bibinfo {author} {\bibfnamefont {F.}~\bibnamefont
  {Ferlaino}},\ }\href {https://doi.org/10.1038/s41586-021-03725-7} {\bibfield
  {journal} {\bibinfo  {journal} {Nature}\ }\textbf {\bibinfo {volume} {596}},\
  \bibinfo {pages} {357} (\bibinfo {year} {2021})}\BibitemShut {NoStop}%
\bibitem [{\citenamefont {Bland}\ \emph
  {et~al.}(2022{\natexlab{a}})\citenamefont {Bland}, \citenamefont {Poli},
  \citenamefont {Politi}, \citenamefont {Klaus}, \citenamefont {Norcia},
  \citenamefont {Ferlaino}, \citenamefont {Santos},\ and\ \citenamefont
  {Bisset}}]{bland_2022_twodimensional}%
  \BibitemOpen
  \bibfield  {author} {\bibinfo {author} {\bibfnamefont {T.}~\bibnamefont
  {Bland}}, \bibinfo {author} {\bibfnamefont {E.}~\bibnamefont {Poli}},
  \bibinfo {author} {\bibfnamefont {C.}~\bibnamefont {Politi}}, \bibinfo
  {author} {\bibfnamefont {L.}~\bibnamefont {Klaus}}, \bibinfo {author}
  {\bibfnamefont {M.~A.}\ \bibnamefont {Norcia}}, \bibinfo {author}
  {\bibfnamefont {F.}~\bibnamefont {Ferlaino}}, \bibinfo {author}
  {\bibfnamefont {L.}~\bibnamefont {Santos}},\ and\ \bibinfo {author}
  {\bibfnamefont {R.~N.}\ \bibnamefont {Bisset}},\ }\href
  {https://doi.org/10.1103/PhysRevLett.128.195302} {\bibfield  {journal}
  {\bibinfo  {journal} {Physical Review Letters}\ }\textbf {\bibinfo {volume}
  {128}},\ \bibinfo {pages} {195302} (\bibinfo {year}
  {2022}{\natexlab{a}})}\BibitemShut {NoStop}%
\bibitem [{\citenamefont {Hertkorn}\ \emph {et~al.}(2021)\citenamefont
  {Hertkorn}, \citenamefont {Schmidt}, \citenamefont {Guo}, \citenamefont
  {B{\"o}ttcher}, \citenamefont {Ng}, \citenamefont {Graham}, \citenamefont
  {Uerlings}, \citenamefont {Langen}, \citenamefont {Zwierlein},\ and\
  \citenamefont {Pfau}}]{hertkorn_2021_pattern}%
  \BibitemOpen
  \bibfield  {author} {\bibinfo {author} {\bibfnamefont {J.}~\bibnamefont
  {Hertkorn}}, \bibinfo {author} {\bibfnamefont {J.-N.}\ \bibnamefont
  {Schmidt}}, \bibinfo {author} {\bibfnamefont {M.}~\bibnamefont {Guo}},
  \bibinfo {author} {\bibfnamefont {F.}~\bibnamefont {B{\"o}ttcher}}, \bibinfo
  {author} {\bibfnamefont {K.~S.~H.}\ \bibnamefont {Ng}}, \bibinfo {author}
  {\bibfnamefont {S.~D.}\ \bibnamefont {Graham}}, \bibinfo {author}
  {\bibfnamefont {P.}~\bibnamefont {Uerlings}}, \bibinfo {author}
  {\bibfnamefont {T.}~\bibnamefont {Langen}}, \bibinfo {author} {\bibfnamefont
  {M.}~\bibnamefont {Zwierlein}},\ and\ \bibinfo {author} {\bibfnamefont
  {T.}~\bibnamefont {Pfau}},\ }\href
  {https://doi.org/10.1103/PhysRevResearch.3.033125} {\bibfield  {journal}
  {\bibinfo  {journal} {Physical Review Research}\ }\textbf {\bibinfo {volume}
  {3}},\ \bibinfo {pages} {033125} (\bibinfo {year} {2021})}\BibitemShut
  {NoStop}%
\bibitem [{\citenamefont {Zhang}\ \emph {et~al.}(2021)\citenamefont {Zhang},
  \citenamefont {Pohl},\ and\ \citenamefont {Maucher}}]{zhang_2021_phases}%
  \BibitemOpen
  \bibfield  {author} {\bibinfo {author} {\bibfnamefont {Y.-C.}\ \bibnamefont
  {Zhang}}, \bibinfo {author} {\bibfnamefont {T.}~\bibnamefont {Pohl}},\ and\
  \bibinfo {author} {\bibfnamefont {F.}~\bibnamefont {Maucher}},\ }\href
  {https://doi.org/10.1103/PhysRevA.104.013310} {\bibfield  {journal} {\bibinfo
   {journal} {Physical Review A}\ }\textbf {\bibinfo {volume} {104}},\ \bibinfo
  {pages} {013310} (\bibinfo {year} {2021})}\BibitemShut {NoStop}%
\bibitem [{\citenamefont {Poli}\ \emph {et~al.}(2021)\citenamefont {Poli},
  \citenamefont {Bland}, \citenamefont {Politi}, \citenamefont {Klaus},
  \citenamefont {Norcia}, \citenamefont {Ferlaino}, \citenamefont {Bisset},\
  and\ \citenamefont {Santos}}]{poli_2021_maintaining}%
  \BibitemOpen
  \bibfield  {author} {\bibinfo {author} {\bibfnamefont {E.}~\bibnamefont
  {Poli}}, \bibinfo {author} {\bibfnamefont {T.}~\bibnamefont {Bland}},
  \bibinfo {author} {\bibfnamefont {C.}~\bibnamefont {Politi}}, \bibinfo
  {author} {\bibfnamefont {L.}~\bibnamefont {Klaus}}, \bibinfo {author}
  {\bibfnamefont {M.~A.}\ \bibnamefont {Norcia}}, \bibinfo {author}
  {\bibfnamefont {F.}~\bibnamefont {Ferlaino}}, \bibinfo {author}
  {\bibfnamefont {R.~N.}\ \bibnamefont {Bisset}},\ and\ \bibinfo {author}
  {\bibfnamefont {L.}~\bibnamefont {Santos}},\ }\href
  {https://doi.org/10.1103/PhysRevA.104.063307} {\bibfield  {journal} {\bibinfo
   {journal} {Physical Review A}\ }\textbf {\bibinfo {volume} {104}},\ \bibinfo
  {pages} {063307} (\bibinfo {year} {2021})}\BibitemShut {NoStop}%
\bibitem [{\citenamefont {Gallem{\'i}}\ and\ \citenamefont
  {Santos}(2022)}]{gallemi_2022_superfluid}%
  \BibitemOpen
  \bibfield  {author} {\bibinfo {author} {\bibfnamefont {A.}~\bibnamefont
  {Gallem{\'i}}}\ and\ \bibinfo {author} {\bibfnamefont {L.}~\bibnamefont
  {Santos}},\ }\href {https://doi.org/10.1103/PhysRevA.106.063301} {\bibfield
  {journal} {\bibinfo  {journal} {Physical Review A}\ }\textbf {\bibinfo
  {volume} {106}},\ \bibinfo {pages} {063301} (\bibinfo {year}
  {2022})}\BibitemShut {NoStop}%
\bibitem [{\citenamefont {Arazo}\ \emph {et~al.}(2023)\citenamefont {Arazo},
  \citenamefont {Gallem{\'i}}, \citenamefont {Guilleumas}, \citenamefont
  {Mayol},\ and\ \citenamefont {Santos}}]{arazo_2023_selfbound}%
  \BibitemOpen
  \bibfield  {author} {\bibinfo {author} {\bibfnamefont {M.}~\bibnamefont
  {Arazo}}, \bibinfo {author} {\bibfnamefont {A.}~\bibnamefont {Gallem{\'i}}},
  \bibinfo {author} {\bibfnamefont {M.}~\bibnamefont {Guilleumas}}, \bibinfo
  {author} {\bibfnamefont {R.}~\bibnamefont {Mayol}},\ and\ \bibinfo {author}
  {\bibfnamefont {L.}~\bibnamefont {Santos}},\ }\href@noop {} {\bibinfo {title}
  {Self-bound crystals of antiparallel dipolar mixtures}} (\bibinfo {year}
  {2023}),\ \Eprint {https://arxiv.org/abs/2303.02087} {arxiv:2303.02087
  [cond-mat]} \BibitemShut {NoStop}%
\bibitem [{\citenamefont {Chomaz}\ \emph {et~al.}(2022)\citenamefont {Chomaz},
  \citenamefont {{Ferrier-Barbut}}, \citenamefont {Ferlaino}, \citenamefont
  {{Laburthe-Tolra}}, \citenamefont {Lev},\ and\ \citenamefont
  {Pfau}}]{chomaz_2022_dipolar}%
  \BibitemOpen
  \bibfield  {author} {\bibinfo {author} {\bibfnamefont {L.}~\bibnamefont
  {Chomaz}}, \bibinfo {author} {\bibfnamefont {I.}~\bibnamefont
  {{Ferrier-Barbut}}}, \bibinfo {author} {\bibfnamefont {F.}~\bibnamefont
  {Ferlaino}}, \bibinfo {author} {\bibfnamefont {B.}~\bibnamefont
  {{Laburthe-Tolra}}}, \bibinfo {author} {\bibfnamefont {B.~L.}\ \bibnamefont
  {Lev}},\ and\ \bibinfo {author} {\bibfnamefont {T.}~\bibnamefont {Pfau}},\
  }\href {https://doi.org/10.1088/1361-6633/aca814} {\bibfield  {journal}
  {\bibinfo  {journal} {Reports on Progress in Physics}\ }\textbf {\bibinfo
  {volume} {86}},\ \bibinfo {pages} {026401} (\bibinfo {year}
  {2022})}\BibitemShut {NoStop}%
\bibitem [{\citenamefont {Sohmen}\ \emph {et~al.}(2021)\citenamefont {Sohmen},
  \citenamefont {Politi}, \citenamefont {Klaus}, \citenamefont {Chomaz},
  \citenamefont {Mark}, \citenamefont {Norcia},\ and\ \citenamefont
  {Ferlaino}}]{sohmen_2021_birth}%
  \BibitemOpen
  \bibfield  {author} {\bibinfo {author} {\bibfnamefont {M.}~\bibnamefont
  {Sohmen}}, \bibinfo {author} {\bibfnamefont {C.}~\bibnamefont {Politi}},
  \bibinfo {author} {\bibfnamefont {L.}~\bibnamefont {Klaus}}, \bibinfo
  {author} {\bibfnamefont {L.}~\bibnamefont {Chomaz}}, \bibinfo {author}
  {\bibfnamefont {M.~J.}\ \bibnamefont {Mark}}, \bibinfo {author}
  {\bibfnamefont {M.~A.}\ \bibnamefont {Norcia}},\ and\ \bibinfo {author}
  {\bibfnamefont {F.}~\bibnamefont {Ferlaino}},\ }\href
  {https://doi.org/10.1103/PhysRevLett.126.233401} {\bibfield  {journal}
  {\bibinfo  {journal} {Physical Review Letters}\ }\textbf {\bibinfo {volume}
  {126}},\ \bibinfo {pages} {233401} (\bibinfo {year} {2021})}\BibitemShut
  {NoStop}%
\bibitem [{\citenamefont {Prasad}\ \emph {et~al.}(2019)\citenamefont {Prasad},
  \citenamefont {Bland}, \citenamefont {Mulkerin}, \citenamefont {Parker},\
  and\ \citenamefont {Martin}}]{prasad_2019_vortex}%
  \BibitemOpen
  \bibfield  {author} {\bibinfo {author} {\bibfnamefont {S.~B.}\ \bibnamefont
  {Prasad}}, \bibinfo {author} {\bibfnamefont {T.}~\bibnamefont {Bland}},
  \bibinfo {author} {\bibfnamefont {B.~C.}\ \bibnamefont {Mulkerin}}, \bibinfo
  {author} {\bibfnamefont {N.~G.}\ \bibnamefont {Parker}},\ and\ \bibinfo
  {author} {\bibfnamefont {A.~M.}\ \bibnamefont {Martin}},\ }\href
  {https://doi.org/10.1103/PhysRevA.100.023625} {\bibfield  {journal} {\bibinfo
   {journal} {Physical Review A}\ }\textbf {\bibinfo {volume} {100}},\ \bibinfo
  {pages} {023625} (\bibinfo {year} {2019})}\BibitemShut {NoStop}%
\bibitem [{\citenamefont {Roccuzzo}\ \emph {et~al.}(2020)\citenamefont
  {Roccuzzo}, \citenamefont {Gallem{\'i}}, \citenamefont {Recati},\ and\
  \citenamefont {Stringari}}]{roccuzzo_2020_rotating}%
  \BibitemOpen
  \bibfield  {author} {\bibinfo {author} {\bibfnamefont {S.~M.}\ \bibnamefont
  {Roccuzzo}}, \bibinfo {author} {\bibfnamefont {A.}~\bibnamefont
  {Gallem{\'i}}}, \bibinfo {author} {\bibfnamefont {A.}~\bibnamefont
  {Recati}},\ and\ \bibinfo {author} {\bibfnamefont {S.}~\bibnamefont
  {Stringari}},\ }\href {https://doi.org/10.1103/PhysRevLett.124.045702}
  {\bibfield  {journal} {\bibinfo  {journal} {Physical Review Letters}\
  }\textbf {\bibinfo {volume} {124}},\ \bibinfo {pages} {045702} (\bibinfo
  {year} {2020})}\BibitemShut {NoStop}%
\bibitem [{\citenamefont {Gallem{\`i}}\ \emph {et~al.}(2020)\citenamefont
  {Gallem{\`i}}, \citenamefont {Roccuzzo}, \citenamefont {Stringari},\ and\
  \citenamefont {Recati}}]{gallemi_2020_quantized}%
  \BibitemOpen
  \bibfield  {author} {\bibinfo {author} {\bibfnamefont {A.}~\bibnamefont
  {Gallem{\`i}}}, \bibinfo {author} {\bibfnamefont {S.~M.}\ \bibnamefont
  {Roccuzzo}}, \bibinfo {author} {\bibfnamefont {S.}~\bibnamefont
  {Stringari}},\ and\ \bibinfo {author} {\bibfnamefont {A.}~\bibnamefont
  {Recati}},\ }\href {https://doi.org/10.1103/PhysRevA.102.023322} {\bibfield
  {journal} {\bibinfo  {journal} {Physical Review A}\ }\textbf {\bibinfo
  {volume} {102}},\ \bibinfo {pages} {023322} (\bibinfo {year}
  {2020})}\BibitemShut {NoStop}%
\bibitem [{\citenamefont {{\v S}indik}\ \emph {et~al.}(2022)\citenamefont {{\v
  S}indik}, \citenamefont {Recati}, \citenamefont {Roccuzzo}, \citenamefont
  {Santos},\ and\ \citenamefont {Stringari}}]{sindik_2022_creation}%
  \BibitemOpen
  \bibfield  {author} {\bibinfo {author} {\bibfnamefont {M.}~\bibnamefont {{\v
  S}indik}}, \bibinfo {author} {\bibfnamefont {A.}~\bibnamefont {Recati}},
  \bibinfo {author} {\bibfnamefont {S.~M.}\ \bibnamefont {Roccuzzo}}, \bibinfo
  {author} {\bibfnamefont {L.}~\bibnamefont {Santos}},\ and\ \bibinfo {author}
  {\bibfnamefont {S.}~\bibnamefont {Stringari}},\ }\href
  {https://doi.org/10.1103/PhysRevA.106.L061303} {\bibfield  {journal}
  {\bibinfo  {journal} {Physical Review A}\ }\textbf {\bibinfo {volume}
  {106}},\ \bibinfo {pages} {L061303} (\bibinfo {year} {2022})}\BibitemShut
  {NoStop}%
\bibitem [{\citenamefont {Klaus}\ \emph {et~al.}(2022)\citenamefont {Klaus},
  \citenamefont {Bland}, \citenamefont {Poli}, \citenamefont {Politi},
  \citenamefont {Lamporesi}, \citenamefont {Casotti}, \citenamefont {Bisset},
  \citenamefont {Mark},\ and\ \citenamefont
  {Ferlaino}}]{klaus_2022_observationa}%
  \BibitemOpen
  \bibfield  {author} {\bibinfo {author} {\bibfnamefont {L.}~\bibnamefont
  {Klaus}}, \bibinfo {author} {\bibfnamefont {T.}~\bibnamefont {Bland}},
  \bibinfo {author} {\bibfnamefont {E.}~\bibnamefont {Poli}}, \bibinfo {author}
  {\bibfnamefont {C.}~\bibnamefont {Politi}}, \bibinfo {author} {\bibfnamefont
  {G.}~\bibnamefont {Lamporesi}}, \bibinfo {author} {\bibfnamefont
  {E.}~\bibnamefont {Casotti}}, \bibinfo {author} {\bibfnamefont {R.~N.}\
  \bibnamefont {Bisset}}, \bibinfo {author} {\bibfnamefont {M.~J.}\
  \bibnamefont {Mark}},\ and\ \bibinfo {author} {\bibfnamefont
  {F.}~\bibnamefont {Ferlaino}},\ }\href
  {https://doi.org/10.1038/s41567-022-01793-8} {\bibfield  {journal} {\bibinfo
  {journal} {Nature Physics}\ }\textbf {\bibinfo {volume} {18}},\ \bibinfo
  {pages} {1453} (\bibinfo {year} {2022})}\BibitemShut {NoStop}%
\bibitem [{\citenamefont {Halder}\ \emph {et~al.}(2022)\citenamefont {Halder},
  \citenamefont {Mukherjee}, \citenamefont {Mistakidis}, \citenamefont {Das},
  \citenamefont {Kevrekidis}, \citenamefont {Panigrahi}, \citenamefont
  {Majumder},\ and\ \citenamefont {Sadeghpour}}]{halder_2022_controla}%
  \BibitemOpen
  \bibfield  {author} {\bibinfo {author} {\bibfnamefont {S.}~\bibnamefont
  {Halder}}, \bibinfo {author} {\bibfnamefont {K.}~\bibnamefont {Mukherjee}},
  \bibinfo {author} {\bibfnamefont {S.~I.}\ \bibnamefont {Mistakidis}},
  \bibinfo {author} {\bibfnamefont {S.}~\bibnamefont {Das}}, \bibinfo {author}
  {\bibfnamefont {P.~G.}\ \bibnamefont {Kevrekidis}}, \bibinfo {author}
  {\bibfnamefont {P.~K.}\ \bibnamefont {Panigrahi}}, \bibinfo {author}
  {\bibfnamefont {S.}~\bibnamefont {Majumder}},\ and\ \bibinfo {author}
  {\bibfnamefont {H.~R.}\ \bibnamefont {Sadeghpour}},\ }\href
  {https://doi.org/10.1103/PhysRevResearch.4.043124} {\bibfield  {journal}
  {\bibinfo  {journal} {Physical Review Research}\ }\textbf {\bibinfo {volume}
  {4}},\ \bibinfo {pages} {043124} (\bibinfo {year} {2022})}\BibitemShut
  {NoStop}%
\bibitem [{\citenamefont {Kirkby}\ \emph {et~al.}(2023)\citenamefont {Kirkby},
  \citenamefont {Bland}, \citenamefont {Ferlaino},\ and\ \citenamefont
  {Bisset}}]{kirkby_2023_spin}%
  \BibitemOpen
  \bibfield  {author} {\bibinfo {author} {\bibfnamefont {W.}~\bibnamefont
  {Kirkby}}, \bibinfo {author} {\bibfnamefont {T.}~\bibnamefont {Bland}},
  \bibinfo {author} {\bibfnamefont {F.}~\bibnamefont {Ferlaino}},\ and\
  \bibinfo {author} {\bibfnamefont {R.~N.}\ \bibnamefont {Bisset}},\
  }\href@noop {} {\bibinfo {title} {Spin rotons and supersolids in binary
  antidipolar condensates}} (\bibinfo {year} {2023}),\ \Eprint
  {https://arxiv.org/abs/2301.08007} {arXiv:2301.08007 [cond-mat.quant-gas]}
  \BibitemShut {NoStop}%
\bibitem [{\citenamefont {Mukherjee}\ \emph {et~al.}(2023)\citenamefont
  {Mukherjee}, \citenamefont {Tengstrand}, \citenamefont {Cardinale},\ and\
  \citenamefont {Reimann}}]{mukherjee_2023_supersolid}%
  \BibitemOpen
  \bibfield  {author} {\bibinfo {author} {\bibfnamefont {K.}~\bibnamefont
  {Mukherjee}}, \bibinfo {author} {\bibfnamefont {M.~N.}\ \bibnamefont
  {Tengstrand}}, \bibinfo {author} {\bibfnamefont {T.~A.}\ \bibnamefont
  {Cardinale}},\ and\ \bibinfo {author} {\bibfnamefont {S.~M.}\ \bibnamefont
  {Reimann}},\ }\href {https://doi.org/10.1103/PhysRevA.108.023302} {\bibfield
  {journal} {\bibinfo  {journal} {Physical Review A}\ }\textbf {\bibinfo
  {volume} {108}},\ \bibinfo {pages} {023302} (\bibinfo {year}
  {2023})}\BibitemShut {NoStop}%
\bibitem [{\citenamefont {Semeghini}\ \emph {et~al.}(2018)\citenamefont
  {Semeghini}, \citenamefont {Ferioli}, \citenamefont {Masi}, \citenamefont
  {Mazzinghi}, \citenamefont {Wolswijk}, \citenamefont {Minardi}, \citenamefont
  {Modugno}, \citenamefont {Modugno}, \citenamefont {Inguscio},\ and\
  \citenamefont {Fattori}}]{semeghini_2018_selfbound}%
  \BibitemOpen
  \bibfield  {author} {\bibinfo {author} {\bibfnamefont {G.}~\bibnamefont
  {Semeghini}}, \bibinfo {author} {\bibfnamefont {G.}~\bibnamefont {Ferioli}},
  \bibinfo {author} {\bibfnamefont {L.}~\bibnamefont {Masi}}, \bibinfo {author}
  {\bibfnamefont {C.}~\bibnamefont {Mazzinghi}}, \bibinfo {author}
  {\bibfnamefont {L.}~\bibnamefont {Wolswijk}}, \bibinfo {author}
  {\bibfnamefont {F.}~\bibnamefont {Minardi}}, \bibinfo {author} {\bibfnamefont
  {M.}~\bibnamefont {Modugno}}, \bibinfo {author} {\bibfnamefont
  {G.}~\bibnamefont {Modugno}}, \bibinfo {author} {\bibfnamefont
  {M.}~\bibnamefont {Inguscio}},\ and\ \bibinfo {author} {\bibfnamefont
  {M.}~\bibnamefont {Fattori}},\ }\href
  {https://doi.org/10.1103/PhysRevLett.120.235301} {\bibfield  {journal}
  {\bibinfo  {journal} {Physical Review Letters}\ }\textbf {\bibinfo {volume}
  {120}},\ \bibinfo {pages} {235301} (\bibinfo {year} {2018})}\BibitemShut
  {NoStop}%
\bibitem [{\citenamefont {Cheiney}\ \emph {et~al.}(2018)\citenamefont
  {Cheiney}, \citenamefont {Cabrera}, \citenamefont {Sanz}, \citenamefont
  {Naylor}, \citenamefont {Tanzi},\ and\ \citenamefont
  {Tarruell}}]{cheiney_2018_bright}%
  \BibitemOpen
  \bibfield  {author} {\bibinfo {author} {\bibfnamefont {P.}~\bibnamefont
  {Cheiney}}, \bibinfo {author} {\bibfnamefont {C.~R.}\ \bibnamefont
  {Cabrera}}, \bibinfo {author} {\bibfnamefont {J.}~\bibnamefont {Sanz}},
  \bibinfo {author} {\bibfnamefont {B.}~\bibnamefont {Naylor}}, \bibinfo
  {author} {\bibfnamefont {L.}~\bibnamefont {Tanzi}},\ and\ \bibinfo {author}
  {\bibfnamefont {L.}~\bibnamefont {Tarruell}},\ }\href
  {https://doi.org/10.1103/PhysRevLett.120.135301} {\bibfield  {journal}
  {\bibinfo  {journal} {Physical Review Letters}\ }\textbf {\bibinfo {volume}
  {120}},\ \bibinfo {pages} {135301} (\bibinfo {year} {2018})}\BibitemShut
  {NoStop}%
\bibitem [{\citenamefont {Ferioli}\ \emph {et~al.}(2019)\citenamefont
  {Ferioli}, \citenamefont {Semeghini}, \citenamefont {Masi}, \citenamefont
  {Giusti}, \citenamefont {Modugno}, \citenamefont {Inguscio}, \citenamefont
  {Gallem{\'i}}, \citenamefont {Recati},\ and\ \citenamefont
  {Fattori}}]{ferioli_2019_collisions}%
  \BibitemOpen
  \bibfield  {author} {\bibinfo {author} {\bibfnamefont {G.}~\bibnamefont
  {Ferioli}}, \bibinfo {author} {\bibfnamefont {G.}~\bibnamefont {Semeghini}},
  \bibinfo {author} {\bibfnamefont {L.}~\bibnamefont {Masi}}, \bibinfo {author}
  {\bibfnamefont {G.}~\bibnamefont {Giusti}}, \bibinfo {author} {\bibfnamefont
  {G.}~\bibnamefont {Modugno}}, \bibinfo {author} {\bibfnamefont
  {M.}~\bibnamefont {Inguscio}}, \bibinfo {author} {\bibfnamefont
  {A.}~\bibnamefont {Gallem{\'i}}}, \bibinfo {author} {\bibfnamefont
  {A.}~\bibnamefont {Recati}},\ and\ \bibinfo {author} {\bibfnamefont
  {M.}~\bibnamefont {Fattori}},\ }\href
  {https://doi.org/10.1103/PhysRevLett.122.090401} {\bibfield  {journal}
  {\bibinfo  {journal} {Physical Review Letters}\ }\textbf {\bibinfo {volume}
  {122}},\ \bibinfo {pages} {090401} (\bibinfo {year} {2019})}\BibitemShut
  {NoStop}%
\bibitem [{\citenamefont {Flynn}\ \emph {et~al.}(2023)\citenamefont {Flynn},
  \citenamefont {Parisi}, \citenamefont {Billam},\ and\ \citenamefont
  {Parker}}]{flynn_2022_quantum}%
  \BibitemOpen
  \bibfield  {author} {\bibinfo {author} {\bibfnamefont {T.~A.}\ \bibnamefont
  {Flynn}}, \bibinfo {author} {\bibfnamefont {L.}~\bibnamefont {Parisi}},
  \bibinfo {author} {\bibfnamefont {T.~P.}\ \bibnamefont {Billam}},\ and\
  \bibinfo {author} {\bibfnamefont {N.~G.}\ \bibnamefont {Parker}},\ }\href
  {https://doi.org/10.1103/PhysRevResearch.5.033167} {\bibfield  {journal}
  {\bibinfo  {journal} {Phys. Rev. Res.}\ }\textbf {\bibinfo {volume} {5}},\
  \bibinfo {pages} {033167} (\bibinfo {year} {2023})}\BibitemShut {NoStop}%
\bibitem [{\citenamefont {Guo}\ \emph {et~al.}(2021)\citenamefont {Guo},
  \citenamefont {Jia}, \citenamefont {Li}, \citenamefont {Ma}, \citenamefont
  {Hutson}, \citenamefont {Cui},\ and\ \citenamefont
  {Wang}}]{guo_2021_leehuangyang}%
  \BibitemOpen
  \bibfield  {author} {\bibinfo {author} {\bibfnamefont {Z.}~\bibnamefont
  {Guo}}, \bibinfo {author} {\bibfnamefont {F.}~\bibnamefont {Jia}}, \bibinfo
  {author} {\bibfnamefont {L.}~\bibnamefont {Li}}, \bibinfo {author}
  {\bibfnamefont {Y.}~\bibnamefont {Ma}}, \bibinfo {author} {\bibfnamefont
  {J.~M.}\ \bibnamefont {Hutson}}, \bibinfo {author} {\bibfnamefont
  {X.}~\bibnamefont {Cui}},\ and\ \bibinfo {author} {\bibfnamefont
  {D.}~\bibnamefont {Wang}},\ }\href
  {https://doi.org/10.1103/PhysRevResearch.3.033247} {\bibfield  {journal}
  {\bibinfo  {journal} {Physical Review Research}\ }\textbf {\bibinfo {volume}
  {3}},\ \bibinfo {pages} {033247} (\bibinfo {year} {2021})}\BibitemShut
  {NoStop}%
\bibitem [{\citenamefont {D'Errico}\ \emph {et~al.}(2019)\citenamefont
  {D'Errico}, \citenamefont {Burchianti}, \citenamefont {Prevedelli},
  \citenamefont {Salasnich}, \citenamefont {Ancilotto}, \citenamefont
  {Modugno}, \citenamefont {Minardi},\ and\ \citenamefont
  {Fort}}]{derrico_2019_observation}%
  \BibitemOpen
  \bibfield  {author} {\bibinfo {author} {\bibfnamefont {C.}~\bibnamefont
  {D'Errico}}, \bibinfo {author} {\bibfnamefont {A.}~\bibnamefont
  {Burchianti}}, \bibinfo {author} {\bibfnamefont {M.}~\bibnamefont
  {Prevedelli}}, \bibinfo {author} {\bibfnamefont {L.}~\bibnamefont
  {Salasnich}}, \bibinfo {author} {\bibfnamefont {F.}~\bibnamefont
  {Ancilotto}}, \bibinfo {author} {\bibfnamefont {M.}~\bibnamefont {Modugno}},
  \bibinfo {author} {\bibfnamefont {F.}~\bibnamefont {Minardi}},\ and\ \bibinfo
  {author} {\bibfnamefont {C.}~\bibnamefont {Fort}},\ }\href
  {https://doi.org/10.1103/PhysRevResearch.1.033155} {\bibfield  {journal}
  {\bibinfo  {journal} {Physical Review Research}\ }\textbf {\bibinfo {volume}
  {1}},\ \bibinfo {pages} {033155} (\bibinfo {year} {2019})}\BibitemShut
  {NoStop}%
\bibitem [{\citenamefont {Schmidt}\ \emph {et~al.}(2022)\citenamefont
  {Schmidt}, \citenamefont {Lassabli{\`e}re}, \citenamefont
  {Qu{\'e}m{\'e}ner},\ and\ \citenamefont {Langen}}]{schmidt_2022_selfbound}%
  \BibitemOpen
  \bibfield  {author} {\bibinfo {author} {\bibfnamefont {M.}~\bibnamefont
  {Schmidt}}, \bibinfo {author} {\bibfnamefont {L.}~\bibnamefont
  {Lassabli{\`e}re}}, \bibinfo {author} {\bibfnamefont {G.}~\bibnamefont
  {Qu{\'e}m{\'e}ner}},\ and\ \bibinfo {author} {\bibfnamefont {T.}~\bibnamefont
  {Langen}},\ }\href {https://doi.org/10.1103/PhysRevResearch.4.013235}
  {\bibfield  {journal} {\bibinfo  {journal} {Physical Review Research}\
  }\textbf {\bibinfo {volume} {4}},\ \bibinfo {pages} {013235} (\bibinfo {year}
  {2022})}\BibitemShut {NoStop}%
\bibitem [{\citenamefont {Trautmann}\ \emph {et~al.}(2018)\citenamefont
  {Trautmann}, \citenamefont {Ilzh{\"o}fer}, \citenamefont {Durastante},
  \citenamefont {Politi}, \citenamefont {Sohmen}, \citenamefont {Mark},\ and\
  \citenamefont {Ferlaino}}]{trautmann_2018_dipolar}%
  \BibitemOpen
  \bibfield  {author} {\bibinfo {author} {\bibfnamefont {A.}~\bibnamefont
  {Trautmann}}, \bibinfo {author} {\bibfnamefont {P.}~\bibnamefont
  {Ilzh{\"o}fer}}, \bibinfo {author} {\bibfnamefont {G.}~\bibnamefont
  {Durastante}}, \bibinfo {author} {\bibfnamefont {C.}~\bibnamefont {Politi}},
  \bibinfo {author} {\bibfnamefont {M.}~\bibnamefont {Sohmen}}, \bibinfo
  {author} {\bibfnamefont {M.~J.}\ \bibnamefont {Mark}},\ and\ \bibinfo
  {author} {\bibfnamefont {F.}~\bibnamefont {Ferlaino}},\ }\href
  {https://doi.org/10.1103/PhysRevLett.121.213601} {\bibfield  {journal}
  {\bibinfo  {journal} {Physical Review Letters}\ }\textbf {\bibinfo {volume}
  {121}},\ \bibinfo {pages} {213601} (\bibinfo {year} {2018})}\BibitemShut
  {NoStop}%
\bibitem [{\citenamefont {Durastante}\ \emph {et~al.}(2020)\citenamefont
  {Durastante}, \citenamefont {Politi}, \citenamefont {Sohmen}, \citenamefont
  {Ilzh{\"o}fer}, \citenamefont {Mark}, \citenamefont {Norcia},\ and\
  \citenamefont {Ferlaino}}]{durastante_2020_feshbach}%
  \BibitemOpen
  \bibfield  {author} {\bibinfo {author} {\bibfnamefont {G.}~\bibnamefont
  {Durastante}}, \bibinfo {author} {\bibfnamefont {C.}~\bibnamefont {Politi}},
  \bibinfo {author} {\bibfnamefont {M.}~\bibnamefont {Sohmen}}, \bibinfo
  {author} {\bibfnamefont {P.}~\bibnamefont {Ilzh{\"o}fer}}, \bibinfo {author}
  {\bibfnamefont {M.~J.}\ \bibnamefont {Mark}}, \bibinfo {author}
  {\bibfnamefont {M.~A.}\ \bibnamefont {Norcia}},\ and\ \bibinfo {author}
  {\bibfnamefont {F.}~\bibnamefont {Ferlaino}},\ }\href
  {https://doi.org/10.1103/PhysRevA.102.033330} {\bibfield  {journal} {\bibinfo
   {journal} {Physical Review A}\ }\textbf {\bibinfo {volume} {102}},\ \bibinfo
  {pages} {033330} (\bibinfo {year} {2020})}\BibitemShut {NoStop}%
\bibitem [{\citenamefont {Politi}\ \emph {et~al.}(2022)\citenamefont {Politi},
  \citenamefont {Trautmann}, \citenamefont {Ilzh{\"o}fer}, \citenamefont
  {Durastante}, \citenamefont {Mark}, \citenamefont {Modugno},\ and\
  \citenamefont {Ferlaino}}]{politi_2022_interspecies}%
  \BibitemOpen
  \bibfield  {author} {\bibinfo {author} {\bibfnamefont {C.}~\bibnamefont
  {Politi}}, \bibinfo {author} {\bibfnamefont {A.}~\bibnamefont {Trautmann}},
  \bibinfo {author} {\bibfnamefont {P.}~\bibnamefont {Ilzh{\"o}fer}}, \bibinfo
  {author} {\bibfnamefont {G.}~\bibnamefont {Durastante}}, \bibinfo {author}
  {\bibfnamefont {M.~J.}\ \bibnamefont {Mark}}, \bibinfo {author}
  {\bibfnamefont {M.}~\bibnamefont {Modugno}},\ and\ \bibinfo {author}
  {\bibfnamefont {F.}~\bibnamefont {Ferlaino}},\ }\href
  {https://doi.org/10.1103/PhysRevA.105.023304} {\bibfield  {journal} {\bibinfo
   {journal} {Physical Review A}\ }\textbf {\bibinfo {volume} {105}},\ \bibinfo
  {pages} {023304} (\bibinfo {year} {2022})}\BibitemShut {NoStop}%
\bibitem [{\citenamefont {Bisset}\ \emph {et~al.}(2021)\citenamefont {Bisset},
  \citenamefont {Ardila},\ and\ \citenamefont {Santos}}]{bisset_2021_quantum}%
  \BibitemOpen
  \bibfield  {author} {\bibinfo {author} {\bibfnamefont {R.~N.}\ \bibnamefont
  {Bisset}}, \bibinfo {author} {\bibfnamefont {L.~A.~P.}\ \bibnamefont
  {Ardila}},\ and\ \bibinfo {author} {\bibfnamefont {L.}~\bibnamefont
  {Santos}},\ }\href {https://doi.org/10.1103/PhysRevLett.126.025301}
  {\bibfield  {journal} {\bibinfo  {journal} {Physical Review Letters}\
  }\textbf {\bibinfo {volume} {126}},\ \bibinfo {pages} {025301} (\bibinfo
  {year} {2021})}\BibitemShut {NoStop}%
\bibitem [{\citenamefont {Smith}\ \emph
  {et~al.}(2021{\natexlab{a}})\citenamefont {Smith}, \citenamefont {Baillie},\
  and\ \citenamefont {Blakie}}]{smith_2021_quantum}%
  \BibitemOpen
  \bibfield  {author} {\bibinfo {author} {\bibfnamefont {J.~C.}\ \bibnamefont
  {Smith}}, \bibinfo {author} {\bibfnamefont {D.}~\bibnamefont {Baillie}},\
  and\ \bibinfo {author} {\bibfnamefont {P.~B.}\ \bibnamefont {Blakie}},\
  }\href {https://doi.org/10.1103/PhysRevLett.126.025302} {\bibfield  {journal}
  {\bibinfo  {journal} {Physical Review Letters}\ }\textbf {\bibinfo {volume}
  {126}},\ \bibinfo {pages} {025302} (\bibinfo {year}
  {2021}{\natexlab{a}})}\BibitemShut {NoStop}%
\bibitem [{\citenamefont {Smith}\ \emph
  {et~al.}(2021{\natexlab{b}})\citenamefont {Smith}, \citenamefont {Blakie},\
  and\ \citenamefont {Baillie}}]{smith_2021_approximate}%
  \BibitemOpen
  \bibfield  {author} {\bibinfo {author} {\bibfnamefont {J.~C.}\ \bibnamefont
  {Smith}}, \bibinfo {author} {\bibfnamefont {P.~B.}\ \bibnamefont {Blakie}},\
  and\ \bibinfo {author} {\bibfnamefont {D.}~\bibnamefont {Baillie}},\ }\href
  {https://doi.org/10.1103/PhysRevA.104.053316} {\bibfield  {journal} {\bibinfo
   {journal} {Physical Review A}\ }\textbf {\bibinfo {volume} {104}},\ \bibinfo
  {pages} {053316} (\bibinfo {year} {2021}{\natexlab{b}})}\BibitemShut
  {NoStop}%
\bibitem [{\citenamefont {Lee}\ \emph {et~al.}(2021)\citenamefont {Lee},
  \citenamefont {Baillie}, \citenamefont {Blakie},\ and\ \citenamefont
  {Bisset}}]{lee_2021_miscibility}%
  \BibitemOpen
  \bibfield  {author} {\bibinfo {author} {\bibfnamefont {A.-C.}\ \bibnamefont
  {Lee}}, \bibinfo {author} {\bibfnamefont {D.}~\bibnamefont {Baillie}},
  \bibinfo {author} {\bibfnamefont {P.~B.}\ \bibnamefont {Blakie}},\ and\
  \bibinfo {author} {\bibfnamefont {R.~N.}\ \bibnamefont {Bisset}},\ }\href
  {https://doi.org/10.1103/PhysRevA.103.063301} {\bibfield  {journal} {\bibinfo
   {journal} {Physical Review A}\ }\textbf {\bibinfo {volume} {103}},\ \bibinfo
  {pages} {063301} (\bibinfo {year} {2021})}\BibitemShut {NoStop}%
\bibitem [{\citenamefont {Scheiermann}\ \emph {et~al.}(2023)\citenamefont
  {Scheiermann}, \citenamefont {Ardila}, \citenamefont {Bland}, \citenamefont
  {Bisset},\ and\ \citenamefont {Santos}}]{scheiermann_2023_catalyzation}%
  \BibitemOpen
  \bibfield  {author} {\bibinfo {author} {\bibfnamefont {D.}~\bibnamefont
  {Scheiermann}}, \bibinfo {author} {\bibfnamefont {L.~A.~P.}\ \bibnamefont
  {Ardila}}, \bibinfo {author} {\bibfnamefont {T.}~\bibnamefont {Bland}},
  \bibinfo {author} {\bibfnamefont {R.~N.}\ \bibnamefont {Bisset}},\ and\
  \bibinfo {author} {\bibfnamefont {L.}~\bibnamefont {Santos}},\ }\href
  {https://doi.org/10.1103/PhysRevA.107.L021302} {\bibfield  {journal}
  {\bibinfo  {journal} {Physical Review A}\ }\textbf {\bibinfo {volume}
  {107}},\ \bibinfo {pages} {L021302} (\bibinfo {year} {2023})}\BibitemShut
  {NoStop}%
\bibitem [{\citenamefont {Halder}\ \emph {et~al.}(2023)\citenamefont {Halder},
  \citenamefont {Das},\ and\ \citenamefont
  {Majumder}}]{halder_2023_twodimensional}%
  \BibitemOpen
  \bibfield  {author} {\bibinfo {author} {\bibfnamefont {S.}~\bibnamefont
  {Halder}}, \bibinfo {author} {\bibfnamefont {S.}~\bibnamefont {Das}},\ and\
  \bibinfo {author} {\bibfnamefont {S.}~\bibnamefont {Majumder}},\ }\href
  {https://doi.org/10.1103/PhysRevA.107.063303} {\bibfield  {journal} {\bibinfo
   {journal} {Physical Review A}\ }\textbf {\bibinfo {volume} {107}},\ \bibinfo
  {pages} {063303} (\bibinfo {year} {2023})}\BibitemShut {NoStop}%
\bibitem [{\citenamefont {Li}\ \emph {et~al.}(2022)\citenamefont {Li},
  \citenamefont {Le},\ and\ \citenamefont {Saito}}]{li_2022_longlifetime}%
  \BibitemOpen
  \bibfield  {author} {\bibinfo {author} {\bibfnamefont {S.}~\bibnamefont
  {Li}}, \bibinfo {author} {\bibfnamefont {U.~N.}\ \bibnamefont {Le}},\ and\
  \bibinfo {author} {\bibfnamefont {H.}~\bibnamefont {Saito}},\ }\href
  {https://doi.org/10.1103/PhysRevA.105.L061302} {\bibfield  {journal}
  {\bibinfo  {journal} {Physical Review A}\ }\textbf {\bibinfo {volume}
  {105}},\ \bibinfo {pages} {L061302} (\bibinfo {year} {2022})}\BibitemShut
  {NoStop}%
\bibitem [{\citenamefont {Bland}\ \emph
  {et~al.}(2022{\natexlab{b}})\citenamefont {Bland}, \citenamefont {Poli},
  \citenamefont {Ardila}, \citenamefont {Santos}, \citenamefont {Ferlaino},\
  and\ \citenamefont {Bisset}}]{bland_2022_alternatingdomain}%
  \BibitemOpen
  \bibfield  {author} {\bibinfo {author} {\bibfnamefont {T.}~\bibnamefont
  {Bland}}, \bibinfo {author} {\bibfnamefont {E.}~\bibnamefont {Poli}},
  \bibinfo {author} {\bibfnamefont {L.~A.~P.}\ \bibnamefont {Ardila}}, \bibinfo
  {author} {\bibfnamefont {L.}~\bibnamefont {Santos}}, \bibinfo {author}
  {\bibfnamefont {F.}~\bibnamefont {Ferlaino}},\ and\ \bibinfo {author}
  {\bibfnamefont {R.~N.}\ \bibnamefont {Bisset}},\ }\href
  {https://doi.org/10.1103/PhysRevA.106.053322} {\bibfield  {journal} {\bibinfo
   {journal} {Physical Review A}\ }\textbf {\bibinfo {volume} {106}},\ \bibinfo
  {pages} {053322} (\bibinfo {year} {2022}{\natexlab{b}})}\BibitemShut
  {NoStop}%
\bibitem [{\citenamefont {Chen}\ \emph {et~al.}(2019)\citenamefont {Chen},
  \citenamefont {Li}, \citenamefont {Proukakis},\ and\ \citenamefont
  {Malomed}}]{chen_2019_immiscible}%
  \BibitemOpen
  \bibfield  {author} {\bibinfo {author} {\bibfnamefont {Z.}~\bibnamefont
  {Chen}}, \bibinfo {author} {\bibfnamefont {Y.}~\bibnamefont {Li}}, \bibinfo
  {author} {\bibfnamefont {N.~P.}\ \bibnamefont {Proukakis}},\ and\ \bibinfo
  {author} {\bibfnamefont {B.~A.}\ \bibnamefont {Malomed}},\ }\href
  {https://doi.org/10.1088/1367-2630/ab3207} {\bibfield  {journal} {\bibinfo
  {journal} {New Journal of Physics}\ }\textbf {\bibinfo {volume} {21}},\
  \bibinfo {pages} {073058} (\bibinfo {year} {2019})}\BibitemShut {NoStop}%
\bibitem [{\citenamefont {Bandyopadhyay}\ \emph {et~al.}(2017)\citenamefont
  {Bandyopadhyay}, \citenamefont {Roy},\ and\ \citenamefont
  {Angom}}]{bandyopadhyay_2017_dynamics}%
  \BibitemOpen
  \bibfield  {author} {\bibinfo {author} {\bibfnamefont {S.}~\bibnamefont
  {Bandyopadhyay}}, \bibinfo {author} {\bibfnamefont {A.}~\bibnamefont {Roy}},\
  and\ \bibinfo {author} {\bibfnamefont {D.}~\bibnamefont {Angom}},\ }\href
  {https://doi.org/10.1103/PhysRevA.96.043603} {\bibfield  {journal} {\bibinfo
  {journal} {Phys. Rev. A}\ }\textbf {\bibinfo {volume} {96}},\ \bibinfo
  {pages} {043603} (\bibinfo {year} {2017})}\BibitemShut {NoStop}%
\bibitem [{\citenamefont {Leggett}(1998)}]{leggett_1998_superfluid}%
  \BibitemOpen
  \bibfield  {author} {\bibinfo {author} {\bibfnamefont {A.~J.}\ \bibnamefont
  {Leggett}},\ }\href {https://doi.org/10.1023/B:JOSS.0000033170.38619.6c}
  {\bibfield  {journal} {\bibinfo  {journal} {Journal of Statistical Physics}\
  }\textbf {\bibinfo {volume} {93}},\ \bibinfo {pages} {927} (\bibinfo {year}
  {1998})}\BibitemShut {NoStop}%
\bibitem [{\citenamefont {Chauveau}\ \emph {et~al.}(2023)\citenamefont
  {Chauveau}, \citenamefont {Maury}, \citenamefont {Rabec}, \citenamefont
  {Heintze}, \citenamefont {Brochier}, \citenamefont {Nascimbene},
  \citenamefont {Dalibard}, \citenamefont {Beugnon}, \citenamefont {Roccuzzo},\
  and\ \citenamefont {Stringari}}]{chauveau_2023_superfluid}%
  \BibitemOpen
  \bibfield  {author} {\bibinfo {author} {\bibfnamefont {G.}~\bibnamefont
  {Chauveau}}, \bibinfo {author} {\bibfnamefont {C.}~\bibnamefont {Maury}},
  \bibinfo {author} {\bibfnamefont {F.}~\bibnamefont {Rabec}}, \bibinfo
  {author} {\bibfnamefont {C.}~\bibnamefont {Heintze}}, \bibinfo {author}
  {\bibfnamefont {G.}~\bibnamefont {Brochier}}, \bibinfo {author}
  {\bibfnamefont {S.}~\bibnamefont {Nascimbene}}, \bibinfo {author}
  {\bibfnamefont {J.}~\bibnamefont {Dalibard}}, \bibinfo {author}
  {\bibfnamefont {J.}~\bibnamefont {Beugnon}}, \bibinfo {author} {\bibfnamefont
  {S.~M.}\ \bibnamefont {Roccuzzo}},\ and\ \bibinfo {author} {\bibfnamefont
  {S.}~\bibnamefont {Stringari}},\ }\href
  {https://doi.org/10.1103/PhysRevLett.130.226003} {\bibfield  {journal}
  {\bibinfo  {journal} {Physical Review Letters}\ }\textbf {\bibinfo {volume}
  {130}},\ \bibinfo {pages} {226003} (\bibinfo {year} {2023})}\BibitemShut
  {NoStop}%
\bibitem [{\citenamefont {Diniz}\ \emph {et~al.}(2020)\citenamefont {Diniz},
  \citenamefont {Oliveira}, \citenamefont {Lima},\ and\ \citenamefont
  {Henn}}]{diniz_2020_ground}%
  \BibitemOpen
  \bibfield  {author} {\bibinfo {author} {\bibfnamefont {P.~C.}\ \bibnamefont
  {Diniz}}, \bibinfo {author} {\bibfnamefont {E.~A.~B.}\ \bibnamefont
  {Oliveira}}, \bibinfo {author} {\bibfnamefont {A.~R.~P.}\ \bibnamefont
  {Lima}},\ and\ \bibinfo {author} {\bibfnamefont {E.~A.~L.}\ \bibnamefont
  {Henn}},\ }\href {https://doi.org/10.1038/s41598-020-61657-0} {\bibfield
  {journal} {\bibinfo  {journal} {Scientific Reports}\ }\textbf {\bibinfo
  {volume} {10}},\ \bibinfo {pages} {4831} (\bibinfo {year}
  {2020})}\BibitemShut {NoStop}%
\bibitem [{\citenamefont {Ferrier-Barbut}\ \emph {et~al.}(2014)\citenamefont
  {Ferrier-Barbut}, \citenamefont {Delehaye}, \citenamefont {Laurent},
  \citenamefont {Grier}, \citenamefont {Pierce}, \citenamefont {Rem},
  \citenamefont {Chevy},\ and\ \citenamefont
  {Salomon}}]{ferrier-barbut_2014_mixture}%
  \BibitemOpen
  \bibfield  {author} {\bibinfo {author} {\bibfnamefont {I.}~\bibnamefont
  {Ferrier-Barbut}}, \bibinfo {author} {\bibfnamefont {M.}~\bibnamefont
  {Delehaye}}, \bibinfo {author} {\bibfnamefont {S.}~\bibnamefont {Laurent}},
  \bibinfo {author} {\bibfnamefont {A.~T.}\ \bibnamefont {Grier}}, \bibinfo
  {author} {\bibfnamefont {M.}~\bibnamefont {Pierce}}, \bibinfo {author}
  {\bibfnamefont {B.~S.}\ \bibnamefont {Rem}}, \bibinfo {author} {\bibfnamefont
  {F.}~\bibnamefont {Chevy}},\ and\ \bibinfo {author} {\bibfnamefont
  {C.}~\bibnamefont {Salomon}},\ }\href
  {https://doi.org/10.1126/science.1255380} {\bibfield  {journal} {\bibinfo
  {journal} {Science}\ }\textbf {\bibinfo {volume} {345}},\ \bibinfo {pages}
  {1035} (\bibinfo {year} {2014})},\ \Eprint
  {https://arxiv.org/abs/https://www.science.org/doi/pdf/10.1126/science.1255380}
  {https://www.science.org/doi/pdf/10.1126/science.1255380} \BibitemShut
  {NoStop}%
\bibitem [{\citenamefont {{Ruiz-Tijerina}}(2023)}]{ruiz-tijerina_2023_bose}%
  \BibitemOpen
  \bibfield  {author} {\bibinfo {author} {\bibfnamefont {D.~A.}\ \bibnamefont
  {{Ruiz-Tijerina}}},\ }\href {https://doi.org/10.1038/s41563-022-01469-x}
  {\bibfield  {journal} {\bibinfo  {journal} {Nature Materials}\ }\textbf
  {\bibinfo {volume} {22}},\ \bibinfo {pages} {153} (\bibinfo {year}
  {2023})}\BibitemShut {NoStop}%
\bibitem [{\citenamefont {Ravensbergen}\ \emph {et~al.}(2020)\citenamefont
  {Ravensbergen}, \citenamefont {Soave}, \citenamefont {Corre}, \citenamefont
  {Kreyer}, \citenamefont {Huang}, \citenamefont {Kirilov},\ and\ \citenamefont
  {Grimm}}]{ravensbergen_2020_resonantly}%
  \BibitemOpen
  \bibfield  {author} {\bibinfo {author} {\bibfnamefont {C.}~\bibnamefont
  {Ravensbergen}}, \bibinfo {author} {\bibfnamefont {E.}~\bibnamefont {Soave}},
  \bibinfo {author} {\bibfnamefont {V.}~\bibnamefont {Corre}}, \bibinfo
  {author} {\bibfnamefont {M.}~\bibnamefont {Kreyer}}, \bibinfo {author}
  {\bibfnamefont {B.}~\bibnamefont {Huang}}, \bibinfo {author} {\bibfnamefont
  {E.}~\bibnamefont {Kirilov}},\ and\ \bibinfo {author} {\bibfnamefont
  {R.}~\bibnamefont {Grimm}},\ }\href
  {https://doi.org/10.1103/PhysRevLett.124.203402} {\bibfield  {journal}
  {\bibinfo  {journal} {Phys. Rev. Lett.}\ }\textbf {\bibinfo {volume} {124}},\
  \bibinfo {pages} {203402} (\bibinfo {year} {2020})}\BibitemShut {NoStop}%
\bibitem [{\citenamefont {Saboo}\ \emph {et~al.}(2023)\citenamefont {Saboo},
  \citenamefont {Halder}, \citenamefont {Das},\ and\ \citenamefont
  {Majumder}}]{saboo_2023_rayleightaylor}%
  \BibitemOpen
  \bibfield  {author} {\bibinfo {author} {\bibfnamefont {A.}~\bibnamefont
  {Saboo}}, \bibinfo {author} {\bibfnamefont {S.}~\bibnamefont {Halder}},
  \bibinfo {author} {\bibfnamefont {S.}~\bibnamefont {Das}},\ and\ \bibinfo
  {author} {\bibfnamefont {S.}~\bibnamefont {Majumder}},\ }\href
  {https://doi.org/10.1103/PhysRevA.108.013320} {\bibfield  {journal} {\bibinfo
   {journal} {Phys. Rev. A}\ }\textbf {\bibinfo {volume} {108}},\ \bibinfo
  {pages} {013320} (\bibinfo {year} {2023})}\BibitemShut {NoStop}%
\bibitem [{\citenamefont {Raghunandan}\ \emph {et~al.}(2015)\citenamefont
  {Raghunandan}, \citenamefont {Mishra}, \citenamefont {\L{}akomy},
  \citenamefont {Pedri}, \citenamefont {Santos},\ and\ \citenamefont
  {Nath}}]{raghunandan_2015_twodimensional}%
  \BibitemOpen
  \bibfield  {author} {\bibinfo {author} {\bibfnamefont {M.}~\bibnamefont
  {Raghunandan}}, \bibinfo {author} {\bibfnamefont {C.}~\bibnamefont {Mishra}},
  \bibinfo {author} {\bibfnamefont {K.}~\bibnamefont {\L{}akomy}}, \bibinfo
  {author} {\bibfnamefont {P.}~\bibnamefont {Pedri}}, \bibinfo {author}
  {\bibfnamefont {L.}~\bibnamefont {Santos}},\ and\ \bibinfo {author}
  {\bibfnamefont {R.}~\bibnamefont {Nath}},\ }\href
  {https://doi.org/10.1103/PhysRevA.92.013637} {\bibfield  {journal} {\bibinfo
  {journal} {Phys. Rev. A}\ }\textbf {\bibinfo {volume} {92}},\ \bibinfo
  {pages} {013637} (\bibinfo {year} {2015})}\BibitemShut {NoStop}%
\bibitem [{\citenamefont {Das}\ \emph {et~al.}(2022)\citenamefont {Das},
  \citenamefont {Mukherjee},\ and\ \citenamefont {Majumder}}]{das_2022_vortex}%
  \BibitemOpen
  \bibfield  {author} {\bibinfo {author} {\bibfnamefont {S.}~\bibnamefont
  {Das}}, \bibinfo {author} {\bibfnamefont {K.}~\bibnamefont {Mukherjee}},\
  and\ \bibinfo {author} {\bibfnamefont {S.}~\bibnamefont {Majumder}},\ }\href
  {https://doi.org/10.1103/PhysRevA.106.023306} {\bibfield  {journal} {\bibinfo
   {journal} {Phys. Rev. A}\ }\textbf {\bibinfo {volume} {106}},\ \bibinfo
  {pages} {023306} (\bibinfo {year} {2022})}\BibitemShut {NoStop}%
\bibitem [{\citenamefont {Crank}\ and\ \citenamefont
  {Nicolson}(1947)}]{crank_1947_practical}%
  \BibitemOpen
  \bibfield  {author} {\bibinfo {author} {\bibfnamefont {J.}~\bibnamefont
  {Crank}}\ and\ \bibinfo {author} {\bibfnamefont {P.}~\bibnamefont
  {Nicolson}},\ }\href {https://doi.org/10.1017/S0305004100023197} {\bibfield
  {journal} {\bibinfo  {journal} {Mathematical Proceedings of the Cambridge
  Philosophical Society}\ }\textbf {\bibinfo {volume} {43}},\ \bibinfo {pages}
  {50} (\bibinfo {year} {1947})}\BibitemShut {NoStop}%
\end{thebibliography}%

\end{document}